\documentclass[english,12pt,letter]{article}
\pdfoutput=1

\usepackage{jcappub}

\usepackage[latin1]{inputenc}
\usepackage{amsmath}
\usepackage{amssymb}
\usepackage{graphicx}
\usepackage[caption=false]{subfig}
\usepackage{amsthm}
\usepackage{bm}
\usepackage{color}
\usepackage{placeins}

\makeatletter
\@ifundefined{textcolor}{}
{
 \definecolor{BLACK}{gray}{0}
 \definecolor{WHITE}{gray}{1}
 \definecolor{RED}{rgb}{1,0,0}
 \definecolor{GREEN}{rgb}{0,1,0}
 \definecolor{BLUE}{rgb}{0,0,1}
 \definecolor{CYAN}{cmyk}{1,0,0,0}
 \definecolor{MAGENTA}{cmyk}{0,1,0,0}
 \definecolor{YELLOW}{cmyk}{0,0,1,0}
 }

\usepackage{amsfonts}
\usepackage{dcolumn}

\def\be{\begin{equation}}
\def\ee{\end{equation}}
\def\ba{\begin{eqnarray}}
\def\ea{\end{eqnarray}}
\def\bs{\begin{subequations}}
\def\es{\end{subequations}}

\usepackage{babel}

\title{Exact Evolution of Discrete Relativistic Cosmological Models}

\author[a]{Timothy Clifton,}

\author[b]{Daniele Gregoris,}

\author[b,c]{\\Kjell Rosquist,}

\author[a]{and Reza Tavakol}

\affiliation[a]{School of Physics and Astronomy, Queen Mary University of London, UK.}

\affiliation[b]{Department of Physics, Stockholm University, 106 91
  Stockholm, Sweden.}

\affiliation[c]{ICRANet, Piazza della Repubblica, 10,  I--65122 Pescara, Italy.}

\abstract{We study the effects of inhomogeneities on the evolution of the Universe, by considering  a range of cosmological models with discretized matter content. This is done using exact and fully relativistic methods that exploit the symmetries in and about submanifolds of spacetimes that themselves possess no continuous global symmetries. These methods allow us to follow the evolution of our models throughout their entire history, far beyond what has previously been possible. We find that while some space-like curves collapse to anisotropic singularities in finite time, others remain non-singular forever. The resulting picture is of a cosmological spacetime in which some behaviour remains close to Friedmann-like, while other behaviours deviate radically. In particular, we find that large-scale acceleration is possible without any violation of the energy conditions.}

\arxivnumber{}

\begin{document}

\maketitle

\section{Introduction}

A key question in cosmology at present is how to determine the 
{  effects} of small-scale inhomogeneity on the large-scale evolution of the Universe. In the standard approach to cosmological modelling, one starts by assuming the spacetime to possess symmetries (or algebraic properties) on large enough scales, and then allows small fluctuations around this background as a way of accounting for the structures that astronomers observe. In such an approach the large-scale evolution is therefore put in by hand, rather than being derived systematically. Ideally, we may wish to be able to complement this top-down approach with bottom-up approaches, in which one starts by first specifying the building blocks constituting the Universe,  such as galaxies, clusters of galaxies {\it etc.}, and then {calculates} the large-scale evolution that results from the interaction between these constituent parts. In this way the large-scale expansion can be considered as an emergent phenomenon, rather than as a fixed and immutable background.

The observed Universe is, however, far too complicated to allow a fully realistic construction of this type, as it possesses a great deal of structure on a variety of different scales. Furthermore, the fact that we are limited to making observations along a single past-directed light-cone makes it extremely difficult to even determine which structures exist (see e.g. \cite{lightcone}). Given this complexity, an 
{  important approach in estimating the effects} of inhomogeneity on the large-scale dynamics of the Universe is to proceed by making a detailed step-by-step study of the effects of breaking the assumption of homogeneity in cosmological models in a controlled way. In this connection, a number of studies have recently been made in which the mass in the Universe is taken to be confined to discrete sources, rather than being represented by a continuous fluid that permeates 
{  throughout space}. These studies fall into three broad categories: exact solutions \cite{lattice1}, approximate-analytic approaches \cite{LandW,CandF1,CandF2,CandF3,Clifton,Larena1,Larena2}, and numerical integration of the full Einstein equations \cite{Yoo,BandK8cell,BandK2,BandK3}.

In this paper, as a step {towards} constructing a realistic bottom-up model of the Universe, we employ an exact approach, and consider a sequence of models with an increasing number of discrete sources, taken to be non-rotating, uncharged Schwarzschild-like black holes. These sources are arranged in regular configurations, so that they might be considered as approximately homogeneous when course-grained over sufficiently large scales, 
{  but at the same time allow the effects of small-scale inhomogeneities} on the large-scale evolution of the Universe to be studied in a non-perturbative way. Our central tool in this analysis will be the exploitation of symmetries in and about submanifolds of spacetimes that exhibit no continuous global symmetries whatsoever. This type of approach has 
{  been very successful} when applied to the study of black holes \cite{Misner-63, Brill-Lindquist-63, Gibbons, Cadez}, and more recently was employed by three of us to construct the initial data for relativistic cosmological models that consist of a finite number of discrete masses 
{  regularly arranged} on a closed lattice \cite{lattice1}. The corresponding problem in Newtonian cosmology has been studied in \cite{Newton1, Newton2, Newton3, Newton4, Newton5, Newton6, Newton7, Newton8}.

The goal of the present work is {  to extend our} previous treatment, and study the evolution of relativistic cosmological models with regularly arranged discrete masses. We do this by identifying curves that exhibit local rotational symmetry; a one-dimensional isotropy group that allows the Einstein equations to be integrated exactly. This allows us to calculate 
{  the lengths of the} space-like geodesic curves that connect neighbouring masses, as well as those that correspond to the demarcation lines at the edges of the lattice cells. This is sufficient to allow us to determine a number of interesting properties of the full spacetime, including (i) the fact that the big bang/crunch manifests itself at the marginally trapped surfaces of black holes long before it is ever arrived at in the corresponding Friedmann solution, (ii) the fact  that the edges of cell faces never become singular at any time {during their evolution}, and (iii) the fact that the corners of lattice cells are strongly locally isometric to Minkowski space. They also allow us to consider a number of different measures of the large-scale evolution of these spacetimes, each of which can be compared to Friedmann solutions, and all of which show at least some {deviations from Friedmann-like} behaviour.

An important feature of the methods we employ in this paper is that {they} enable us to incorporate all general relativistic effects in a fully non-perturbative way, 
{  while at the same time allowing the evolution of interesting points and lines in space to be followed}
throughout their entire history. This is far beyond what has so far been possible using other techniques. Moreover, while the primary aim in this paper is to study the cosmological consequences of discretizing the matter content of the Universe, the results we present here also provide potentially interesting insights into the effects of cosmological expansion on otherwise isolated gravitationally bound objects. In particular, we are able to calculate the curvature of spacetime at particular high symmetry points on the marginally trapped surfaces of black holes as a function of cosmological time. This is a quantity that stays constant in stationary black hole solutions, but that can diverge in cosmological models at the big bang/crunch.

In Section II we give a brief account of the orthonormal frame approach that we employ throughout the rest of the paper, as well as the evolution and constraint equations, and the initial data that will be used in the following sections. Section III contains a brief description of the lattice constructions that we use to build our discrete models, and in Section IV we consider curves that exhibit local rotational symmetry. We write down the evolution equations that such curves must satisfy, and go on to give the exact solutions for the evolution of space at every point along them. In Section V we discuss the evolution of the edge lengths for our six different lattice models, together with the corresponding Hubble rates and deceleration parameters. In Section VI we then discuss the evolution of the distance between the horizons of neighbouring masses, both along curves that pass through cell centres, as well as along curves that pass through their corners.  Finally, {  we conclude} in Section VII.

Throughout this work spacetime coordinate indices will be taken to run over the second half of the Greek alphabet ($\mu, \nu, \rho . . . =0-3$), while spatial coordinate indices will be taken to run over the second half of the Latin alphabet ($i,j,k . . . =1-3 $). Orthonormal frame spacetime indices will then be taken to run over the first half of the Latin alphabet ($a,b,c ... =0-3$), while spatial orthonormal frame indices will run over the first half of the Greek alphabet ($\alpha, \beta, \gamma ... = 1-3$).

\section{Orthonormal Frame Approach}

In much of this paper we will use an orthonormal frame approach, and
the notation of van Elst and Uggla \cite{vanElst-Uggla-97}. The first frame vector we
{  consider is} the unit time-like vector $u^{\mu}$, defined such that $u_{\mu} u^{\mu}=-1$. 
The covariant derivative of this vector can be decomposed into irreducible parts such that 
\be
\label{du}
\nabla_{ \mu} u _{\nu} = - u_{\mu} \dot{u}_{\nu} +\theta_{\mu \nu}  =  - u_{\mu} \dot{u}_{\nu} + \sigma_{\mu \nu} + \frac{1}{3} \Theta h_{\mu \nu} - \omega_{\mu \nu},
\ee
where $h_{\mu \nu}=g_{\mu \nu} + u_{\mu } u_{\nu}$ is the projection tensor, and a dot denotes a covariant derivative along $u^{\mu}$, such that $\dot{X} = u^{\mu} \nabla_{\mu} X$. The quantities $\Theta$, $\sigma_{\mu \nu}$ and $\omega_{\mu\nu}$ are the expansion scalar, the shear tensor and the vorticity tensor, respectively. They are the trace, symmetric trace-free and antisymmetric parts of $\theta_{\mu \nu}$, which is the spatial projection of $\nabla_{ \mu} u _{\nu}$. It is convenient to also define
\ba
\sigma^2 &=& \frac{1}{2} \sigma_{\mu \nu} \sigma^{\mu \nu}\\
\omega^2 &=& \frac{1}{2} \omega_{\mu \nu} \omega^{\mu \nu}\\
\omega^{\mu} &=& \frac{1}{2} \eta^{\mu \nu \rho \sigma} \omega_{\nu \rho} u_{\sigma},
\label{du3}
\ea
where $\eta^{\mu \nu \rho \sigma}$ is a totally antisymmetric tensor with 
$\eta^{0123}=(-g)^{-\frac{1}{2}}$, and to decompose the Weyl tensor into 
an electric and a magnetic part {relative to $u^{\mu}$}:
\ba
\label{E}
E_{\mu \nu}  &=& C_{\mu \rho \nu \sigma} u^{\rho} u^{\sigma}\\
H_{\mu \nu}  &=& {}^{\star}C_{\mu \rho \nu \sigma} u^{\rho} u^{\sigma},
\label{H}
\ea
where ${}^{\star}C_{\mu \rho \nu \sigma} = \frac{1}{2} \eta_{\mu \rho}^{\phantom{\mu \rho} \tau \phi} C_{\tau \phi \nu \sigma}$ is the dual of the Weyl tensor. 

To {proceed we need to introduce} three more mutually orthogonal space-like unit vectors, each of which is orthogonal to $u^{\mu}$. We can define such a set of vectors as $\{ e_{\alpha}^{\phantom{\alpha} \mu} \}$, where $\alpha=1-3$ labels each vector individually, and where these vectors are chosen such that $e^{\alpha}_{\phantom{\alpha}\mu} e_{\beta}^{\phantom{\beta}\mu} = \delta^{\alpha}_{\phantom{\alpha} \beta}$ and $e_{\alpha}^{\phantom{\alpha}\mu} u_{\mu}=0$. Relabelling so that $u^{\mu}=e_0^{\phantom{0} \mu}$, and adding this vector to the set containing the other three, then results in a set of four orthonormal frame vectors $\{ e_{a}^{\phantom{a} \mu}\}$, where $a=0-3$. These vectors can be contracted with the components of tensors expressed in a coordinate basis to give them in terms of orthonormal frame indices in the following way:
\be
T^{a b ...}_{\phantom{ a b ...} c d ...} = e^{a}_{\phantom{a} \mu} e^{b}_{\phantom{b} \nu} \cdot\cdot\cdot \; e^{\phantom{c} \rho}_{c} e^{\phantom{d} \sigma}_{d} \cdot\cdot\cdot \; T^{\mu \nu ...}_{\phantom{\mu \nu ...} \rho \sigma ...}.
\ee
The remaining parameters that interest us are then the local angular velocity of the spatial frame vectors,
\be
\label{rotation}
\Omega^{\alpha} = \frac{1}{2} \epsilon^{\alpha \beta \gamma} e^{\phantom{\beta} \mu}_{\beta}  \dot{e}_{\gamma \mu},
\ee
and the spatial commutation functions $\gamma^{\alpha}_{\phantom{\alpha} \beta \gamma}$, defined by
\be
\left[ {\bf e}_{\beta}, {\bf e}_{\gamma} \right] = \gamma^{\alpha}_{\phantom{\alpha} \beta \gamma} {\bf e}_{\alpha}, 
\ee
where ${\bf e}_a = e^{\phantom{a} \mu}_a \partial_{\mu}$. These functions can be decomposed into a 1-index object, $a_{\alpha}$, and a symmetric 2-index object, $n_{\alpha \beta}$, according to \cite{Schucking}
\be
\label{commutation}
\gamma^{\alpha}_{\phantom{\alpha} \beta \gamma} = 2 a_{[\beta} \delta^{\alpha}_{\phantom{\alpha} \gamma]} + \epsilon_{\beta \gamma \delta} n^{\delta \alpha},
\ee
where $\epsilon_{\alpha \beta \gamma}$ is a totally antisymmetric tensor with $\epsilon_{123}=1$. They can also be related to the Ricci rotation coefficients, $\Gamma_{cab} = e_{c i} e_b^{\phantom{b} j} \nabla_j e_a^{\phantom{a} i}$,  by
\be
\gamma^a_{\phantom{a} bc} = -\left( \Gamma^a_{\phantom{a} bc} - \Gamma^a_{\phantom{a} cb} \right), \qquad \Gamma_{abc} = \frac{1}{2} \left( g_{ad} \gamma^d_{\phantom{d} cb} -g_{bd} \gamma^d_{\phantom{d} ca} + g_{cd} \gamma^d_{\phantom{d} ab} \right),
\ee
where $\Gamma_{(ab)c}=0$. The vacuum field equations, Jacobi identities and Bianchi identities can now all be written in terms of the orthonormal frame vectors $\{e^{\phantom{a} \mu}_{a}\}$, the kinematic quantities defined in Eqs. (\ref{du})-(\ref{du3}), the electric and magnetic parts of the Weyl tensor defined in Eqs. (\ref{E}) and (\ref{H}), and the rotation and commutation functions defined in Eqs. (\ref{rotation}) and (\ref{commutation}) \cite{vanElst-Uggla-97}.

\subsection{Evolution and Constraint Equations}

If we choose $\dot{u}^{\mu}=0$ then the evolution equations in vacuum, and 
{  in the absence of} a cosmological constant, can be written as follows \cite{vanElst-Uggla-97}:
\ba
{\bf e}_0 (\Theta) &=& -\frac{1}{3} \Theta^2 - 2 \sigma^2 +2 \omega^2
\label{evo1}\\
{\bf e}_0(\sigma^{\alpha \beta}) &=& - \Theta \sigma^{\alpha \beta} +2 \omega^{(\alpha} \Omega^{\beta)} - {}^{\star}S^{\alpha \beta}- \frac{2}{3} \delta^{\alpha \beta} \omega_{\gamma} \Omega^{\gamma} + 2 \epsilon^{\gamma \delta (\alpha} \Omega_{\gamma} \sigma^{\beta )}_{\phantom{\beta)} \delta}
\label{evo2}\\
{\bf e}_0 (\omega^{\alpha}) &=& -\frac{2}{3} \Theta \omega^{\alpha}+ \sigma^{\alpha}_{\phantom{\alpha} \beta} \omega^{\beta} - \epsilon^{\alpha \beta \gamma} \omega_{\beta} \Omega_{\gamma}
\label{evo3}\\
{\bf e}_0(E^{\alpha \beta}) &=& - \Theta E^{\alpha \beta}+3 \sigma^{(\alpha}_{\phantom{(\alpha} \gamma} E^{\beta) \gamma} +\frac{1}{2} n^{\gamma}_{\phantom{\gamma} \gamma} H^{\alpha \beta} - 3 n^{(\alpha}_{\phantom{(\alpha} \gamma} H^{\beta) \gamma}\label{evo4} \\&& \nonumber  - \delta^{\alpha \beta} \left[ \sigma_{\gamma \delta} E^{\gamma \delta} - n_{\gamma \delta} H^{\gamma\delta} \right] +\epsilon^{\gamma \delta (\alpha} \left[ ({\bf e}_{\gamma} - a_{\gamma}) (H^{\beta)}_{\phantom{\beta)}\delta}) - (\omega_{\gamma} - 2 \Omega_{\gamma} ) E^{\beta)}_{\phantom{\beta)} \delta} \right]
\\
{\bf e}_0(H^{\alpha \beta}) &=& - \Theta H^{\alpha \beta}+3 \sigma^{(\alpha}_{\phantom{(\alpha} \gamma} H^{\beta) \gamma} -\frac{1}{2} n^{\gamma}_{\phantom{\gamma} \gamma} E^{\alpha \beta} + 3 n^{(\alpha}_{\phantom{(\alpha} \gamma} E^{\beta) \gamma} \label{evo5} \\&&  \nonumber - \delta^{\alpha \beta} \left[ \sigma_{\gamma \delta} H^{\gamma \delta} + n_{\gamma \delta} E^{\gamma\delta} \right] -\epsilon^{\gamma \delta (\alpha} \left[ ({\bf e}_{\gamma} - a_{\gamma}) (E^{\beta)}_{\phantom{\beta)}\delta}) + (\omega_{\gamma} - 2 \Omega_{\gamma} ) H^{\beta)}_{\phantom{\beta)} \delta} \right]
\\
{\bf e}_0 (a^{\alpha}) &=& -\frac{1}{3} (\delta^{\alpha \beta} {\bf e}_{\beta}+a^{\alpha}) (\Theta) + \frac{1}{2} ({\bf e}_{\beta} - 2 a_{\beta}) (\sigma^{\alpha \beta})  - \frac{1}{2} \epsilon^{\alpha \beta \gamma} ({\bf e}_{\beta} - 2 a_{\beta}) (\omega_{\gamma} - \Omega_{\gamma})
\qquad \quad
\label{evo6}\\
{\bf e}_0(n^{\alpha \beta}) &=& -\frac{1}{3} \Theta n^{\alpha \beta}- \delta^{\gamma (\alpha} {\bf e}_{\gamma} (\omega^{\beta)} - \Omega^{\beta)}) + 2 \sigma^{(\alpha}_{\phantom{(\alpha} \gamma} n^{\beta ) \gamma} \nonumber\\&&\qquad +\delta^{\alpha \beta} {\bf e}_{\gamma} (\omega^{\gamma} - \Omega^{\gamma}) - \epsilon^{\gamma \delta ( \alpha} \left[ {\bf e}_{\gamma}(\sigma^{\beta)}_{\phantom{\beta)}\delta}) - 2 n^{\beta)}_{\phantom{\beta)}\gamma} (\omega_{\delta}-\Omega_{\delta} ) \right] ,
\label{evo7}
\ea
while the relevant constraints are
\ba
0 &=& -\frac{1}{3} \Theta^2 +\sigma^2 -\omega^2 - 2 \omega_{\alpha} \Omega^{\alpha} - \frac{1}{2} {}^{\star}R
\label{con1}\\
E_{\alpha \beta} &=& \frac{1}{3} \Theta \sigma_{\alpha \beta} - \sigma_{\alpha \gamma} \sigma^{\gamma}_{\phantom{\gamma} \beta} - \omega_{\alpha} \omega_{\beta} - 2 \omega_{(\alpha} \Omega_{\beta)} + \frac{1}{3} \delta_{\alpha \beta} \left[ 2 \sigma^2 +\omega^2 +2 \omega_{\gamma} \Omega^{\gamma} \right] + {}^{\star}S_{\alpha \beta}
\label{con2} \qquad \qquad\\
 H_{\alpha \beta}&=& ({\bf e}_{(\alpha} + a_{(\alpha} ) (\omega_{\beta)}) + \frac{1}{2} n^{\gamma}_{\phantom{\gamma} \gamma} \sigma_{\alpha \beta} - 3 n^{\gamma}_{\phantom{\gamma} (\alpha} \sigma_{\beta) \gamma} - \frac{1}{3} \delta_{\alpha \beta} \left[ ({\bf e}_{\gamma} +a_{\gamma})(\omega^{\gamma}) - 3 n_{\gamma \delta} \sigma^{\gamma \delta} \right] \nonumber \\&&\qquad + \epsilon^{\gamma \delta}_{\phantom{ \gamma \delta}(\alpha} \left[ ({\bf e}_{\vert \gamma \vert} - a_{\vert \gamma \vert} ) (\sigma_{\beta) \delta}) - n_{\beta) \gamma} \omega_{\delta} \right]
\label{con3}\\
0 &=& -\frac{2}{3} \delta ^{\alpha \beta} {\bf e}_{\beta} (\Theta) + ({\bf e}_{\beta} -3 a_{\beta} ) (\sigma^{\alpha \beta}) + n^{\alpha}_{\phantom{\alpha} \beta} \omega^{\beta} - \epsilon^{\alpha \beta \gamma} \left[ ({\bf e}_{\beta} - a_{\beta}) (\omega_{\gamma}) + n_{\beta \delta} \sigma^{\delta}_{\phantom{\delta} \gamma} \right]
\label{con4}\\
0 &=& ({\bf e}_{\beta} - 2 a_{\beta}) (n^{\alpha \beta}) - \frac{2}{3} \Theta \omega^{\alpha} - 2 \sigma^{\alpha}_{\phantom{\alpha} \beta} \omega^{\beta} + \epsilon^{\alpha \beta \gamma} \left[ {\bf e}_{\beta}(a_{\gamma}) + 2 \omega_{\beta} \Omega_{\gamma}\right]
\label{con5}\\
0 &=& ({\bf e}_{\alpha} - 2 a_{\alpha}) (\omega^{\alpha})
\label{con6}\\
0 &=& ({\bf e}_{\beta} -3 a_{\beta})(E^{\alpha \beta}) + 3 \omega_{\beta} H^{\alpha \beta} - \epsilon^{\alpha \beta \gamma} \left[ \sigma_{\beta \delta} H^{\delta}_{\phantom{\delta} \gamma} + n_{\beta \delta} E^{\delta}_{\phantom{\delta} \gamma} \right]
\label{con7}\\
0 &=& ({\bf e}_{\beta} -3 a_{\beta})(H^{\alpha \beta}) - 3 \omega_{\beta} E^{\alpha \beta} + \epsilon^{\alpha \beta \gamma} \left[ \sigma_{\beta \delta} E^{\delta}_{\phantom{\delta} \gamma} - n_{\beta \delta} H^{\delta}_{\phantom{\delta} \gamma} \right] ,
\label{con8}
\ea
where
\ba
{}^{\star}R &=& 2 ( 2 {\bf e}_{\alpha} -3 a_{\alpha}) (a^{\alpha}) - \frac{1}{2} b^{\alpha}_{\phantom{\alpha} \alpha}\\
{}^{\star}S_{\alpha \beta} &=& {\bf e}_{(\alpha} ( a_{\beta)}) + b_{\alpha \beta} - \frac{1}{3} \delta_{\alpha \beta} \left[ {\bf e}_{\gamma} (a^{\gamma}) + b^{\gamma}_{\phantom{\gamma} \gamma} \right] - \epsilon^{\gamma \delta}_{\phantom{\gamma \delta} (\alpha} ({\bf e}_{\vert \gamma \vert} - 2 a_{\vert \gamma \vert} ) (n_{\beta) \delta}) ,
\ea
and $b_{\alpha \beta} = 2 n_{\alpha \gamma} n^{\gamma}_{\phantom{\gamma} \beta} - n^{\gamma}_{\phantom{\gamma} \gamma} n_{\alpha \beta}$. If $\omega^{\alpha}=0$ then ${}^{\star}R$ and ${}^{\star}S_{\alpha \beta}$ are the trace and trace-free parts of the Ricci tensor of the space to which $u^{\mu}$ is orthogonal, respectively. If $\omega^{\alpha} \neq 0$ then {these
quantities} are simply defined by the equations above, and do not correspond to any curvature tensor.

The covariantly defined tensors $\sigma_{\alpha \beta}$, $E_{\alpha \beta}$ and $H_{\alpha \beta}$ all share the properties of being space-like, symmetric and trace-free. They can therefore always be written {in terms of} a set of five variables in the following way \cite{vanElst-Uggla-97}:
\ba
\sigma_{+} &=& - \frac{3}{2} \sigma_{11}\\
\sigma_{-} &=& \frac{\sqrt{3}}{2} (\sigma_{22}-\sigma_{33})
\ea
and
\ba
\sigma_{1} &=& \sqrt{3} \sigma_{23}\\
\sigma_{2} &=& \sqrt{3} \sigma_{31}\\
\sigma_{3} &=& \sqrt{3} \sigma_{12},
\ea
such that $\sigma^2 = \frac{1}{3} [ (\sigma_{+})^2+ (\sigma_{-})^2+ (\sigma_{1})^2+ (\sigma_{2})^2+ (\sigma_{3})^2]$. Similar definitions can be made for $E_{\alpha \beta}$ and $H_{\alpha \beta}$, {\it mutatis mutandis}. In the sections that follow we will make use of this notation, as it simplifies the equations.

\subsection{Initial Data}

We will construct our initial data in the same way as in reference \cite{lattice1}. That is, we will consider a space-like hypersurface about which there exists time-reversal symmetry. Any quantity on this surface that changes sign under the transformation $u^{\mu} \rightarrow - u^{\mu}$ must then vanish. From Eq. (\ref{du}) it can immediately be seen that this includes all of the kinematic quantities:
\be
\sigma_{\mu \nu} =\omega_{\mu \nu}=0= \Theta .
\ee
The spatial rotation vector defined in Eq. (\ref{rotation}) must also vanish, as it contains a derivative along $u^{\mu}$. This is also true of $H_{\mu \nu}$, as $\eta_{\mu \nu \rho\sigma}$ must change sign under $u^{\mu} \rightarrow - u^{\mu}$ if the spatial volume element $\eta_{\mu \nu \rho\sigma} u^{\sigma}$ is to remain invariant. This gives
\be
\Omega^{\alpha} = 0= H_{\mu \nu}.
\ee
Let us now consider the constraint equations that result from this simplification. Under these restrictions Eq. (\ref{con1}) becomes
\be
\label{geo1}
{}^{\star}R=0.
\ee
This is the principal equation of geometrostatics. The vanishing of $\omega^{\mu}$ ensures that there exist 3-spaces that are everywhere orthogonal to $u^{\mu}$. From Eqs. (\ref{con2}) and (\ref{geo1}) it can then be seen that the electric part of the Weyl tensor is identical to the Ricci tensor of these spaces:
\be
\label{Eab}
E_{\alpha \beta} = {}^{\star}R_{\alpha \beta}. 
\ee
Together with the vanishing of $\omega^{\mu}$ and $\sigma^{\mu \nu}$ this result means that Eq. (\ref{con7}) is automatically satisfied by the contracted Bianchi identity on the 3-space.  Eq. (\ref{con5}) gives three constraints on the commutation functions of the three spatial frame vectors, and all other constraint equations can then be seen to vanish identically. It is therefore the case that Eq. (\ref{geo1}) is the only constraint that needs to be solved, as is well known from the study of geometrostatics \cite{Misner-63}.

{The solution to Eq. (\ref{geo1}) on a time-symmetric 3-sphere is already known, and is given by \cite{lattice1}}
\be
\label{init1}
dl^2 = \psi^4 \left( d \chi^2 + \sin^2 \chi d \Omega^2 \right),
\ee
where $d\Omega^2 = d\theta^2 + \sin^2 \theta d \phi^2$, and where $\psi=\psi(\chi, \theta, \phi )$ is given by
\be
\label{init2}
\psi (\chi, \theta, \phi) = \sum_{i=1}^{N} \frac{\sqrt{\tilde{m}_i}}{2 f_i (\chi, \theta, \phi)}.
\ee
The parameters $\tilde{m}_i$ in this expression are a set of constants for each of the $i$ terms that appear in Eq. (\ref{init2}), while the functions $f_i$ are given by $f_i(\chi, \theta, \phi)= \sin (\chi_i/2)$. The variables $\chi_i$ here correspond to a new set of coordinates $(\chi_i, \theta_i, \phi_i)$ that are obtained from rotating the original coordinates $(\chi, \theta, \phi)$ until the $i$'th source appears at the position $\chi_i=0$. It is important to realise that the constants $\tilde{m}_i$ that {appear} in Eq. (\ref{init2}) do not themselves directly correspond to the proper mass of each of the $i$ sources that appear in this equation, but instead must be transformed into this quantity in an appropriate way (see \cite{lattice1} for details). 

As {our aim is} to study the evolution of this initial data, it is interesting to see what form the evolution equations take on the surface of time-symmetry. Eqs. (\ref{evo1}), (\ref{evo2}) and (\ref{evo3}) reduce to
\ba
{\bf e}_0(\Theta) &=& 0\\
{\bf e}_0(\sigma^{\alpha \beta}) &=& -E^{\alpha \beta}\\
{\bf e}_0(\omega^{\alpha}) &=& 0.
\ea
This shows that at the time-symmetric hypersurface, which corresponds to a maximum of expansion in the cosmological context \cite{lattice1}, the first derivative of the expansion scalar must vanish, while the first derivative of the shear tensor will in general be non-zero. This is very different behaviour to the Friedmann solutions, where shear plays no role at all in the evolution, and where ${\bf e}_0(\Theta) \neq 0$ at the maximum of expansion. The last of these equations shows that the vorticity tensor does not have any dynamics at the initial surface, and it can be seen from Eq. (\ref{evo3}) that this is in fact true of the entire evolution: $\omega^{\alpha}=0$ for all time if it is zero initially. This means that $u^{\mu}$ is always hypersurface orthogonal, and that we can write $u_{\mu}=t_{,\mu}$, where $t$ is proper time along the integral curves of $u^{\mu}$.

All other evolution equations vanish {identically}, with the exception of Eq. (\ref{evo5}), which becomes
\ba
\label{evo5b}
{\bf e}_0(H^{\alpha \beta}) &=&-\frac{1}{2} n^{\gamma}_{\phantom{\gamma} \gamma} E^{\alpha \beta} + 3 n^{(\alpha}_{\phantom{(\alpha} \gamma} E^{\beta) \gamma}   \\&&- \delta^{\alpha \beta}  n_{\gamma \delta} E^{\gamma\delta} -\epsilon^{\gamma \delta (\alpha} ({\bf e}_{\gamma} - a_{\gamma}) (E^{\beta)}_{\phantom{\beta)}\delta}).  \nonumber
\ea
From this equation it may initially appear that in general we have ${\bf e}_0(H^{\alpha \beta}) \neq 0$ on a time-symmetric surface. However, it can be shown that in this case the 3-Cotton-York tensor can be written ${}^{\star}C_{\alpha \beta} = - {\bf e}_0(H_{\alpha \beta})$ (see \cite{vanElst-Uggla-97} for the expression in general). As the Cotton-York tensor vanishes in conformally flat spaces, it can be seen that the initial data specified in Eq. (\ref{init1}) results in 
\be
{\bf e}_0(H^{\alpha \beta}) = 0.
\ee 
Eq. (\ref{evo5b}) then just states that $E_{\mu \nu}$ must be curl-free at the moment of time-symmetry (see \cite{Ellis-vanElst-1998} for the definition of curl in this context).

In what follows we will often refer to events close to the time-symmetric surface as having happened at ``early times''. This is because we choose to specify our initial data on this surface, and the reader should not confuse this phrase with a reference to an event that happens close to the big bang. Likewise, by ``late times'' we will mean to refer to times that occur in the far future of our initial data, rather than being far from the big bang.

\section{Lattice Constructions}
\label{sec:lattices}

As already stated, in this paper we will be considering the evolution of the lattices constructed in reference \cite{lattice1}. For completeness, we will briefly summarise these constructions here.

We wish to regularly arrange a finite number of masses on a 3-sphere. The simplest way to achieve this is to consider the six possible convex regular polychora (4-polytopes). These are the 4-dimensional analogues of the Platonic solids, and are the only regular constructions it is 
{  possible to make} from the 3-dimensional polyhedra \cite{Coxeter}. In the same way that each regular convex polyhedron can be used to prescribe a regular tiling of a 2-sphere, each polychoron can be used to provide a regular tiling of a 3-sphere. The polyhedra that make up each polychoron are then described as the ``cells'' of the lattice that this tiling creates, and the polygons that make up the polyhedra are described as the ``faces'' of the cells. We then construct our lattice by placing a mass at the centre of each cell. We will call the boundary curves of each of the polygons the ``edges'' of the lattice cells, and the points at which these edges meet will be called the ``vertices''. Each vertex has associated with it its own figure, which is the shape formed if the vertex was cut off the polychoron. We will find in what follows that the vertices will be very important in the evolution of the edges, and that the vertex figure in particular will signify what type of behaviour is expected. This information is summarised in Table \ref{table1}.

\begin{table*}[t!]
\begin{center}
\begin{tabular}{|c|c|c|c|c|c|}
\hline
$\begin{array}{c}  \bf{Lattice}\\
  \bf{Type}  \end{array}$ 
&  $\begin{array}{c} \textbf{Number}\\
  \textbf{of Faces} \end{array}$
& $\begin{array}{c} \textbf{Number}\\
  \textbf{of Edges} \end{array}$ 
& $\begin{array}{c} \textbf{Number}\\
  \textbf{of Vertices} \end{array}$ 
& $\begin{array}{c} \textbf{Vertex}\\
  \textbf{Figure} \end{array}$ 
& $\begin{array}{c} \textbf{Schl{\" a}fli}\\
  \textbf{Symbols} \end{array}$ \\
\hline
$\begin{array}{c}  {5{\rm -cell}}\\ {({\rm Tetrahedra})}  \end{array}$ & $\begin{array}{c} {10}\\ {({\rm Triangles})} \end{array}$ & $\begin{array}{c} {10}\\ {} \end{array}$ & $\begin{array}{c} {5}\\ {} \end{array}$ & $\begin{array}{c} {{\rm Tetrahedron }}\\ {} \end{array}$ & $\begin{array}{c} {\{333\} } \\ {} \end{array}$ \\
$\begin{array}{c} {8{\rm -cell}}\\ {({\rm Cubes})} \end{array}$ & $\begin{array}{c} {24}\\ {({\rm  Squares})} \end{array}$ & $\begin{array}{c} {32}\\ {} \end{array}$ & $\begin{array}{c} {16}\\ {} \end{array}$ & $\begin{array}{c} {{\rm Tetrahedron }}\\ {} \end{array}$ & $\begin{array}{c} {\{433\} }\\ {} \end{array}$  \\
$\begin{array}{c} {16{\rm -cell}}\\ {({\rm Tetrahedra})} \end{array}$ & $\begin{array}{c} {32}\\ {({\rm Triangles})} \end{array}$ & $\begin{array}{c} {24}\\ {} \end{array}$ & $\begin{array}{c} {8}\\ {} \end{array}$ & $\begin{array}{c} {{\rm Octahedron }}\\ {} \end{array}$ & $\begin{array}{c} {\{334\} }\\ {} \end{array}$ \\
$\begin{array}{c} {24{\rm -cell}}\\ {({\rm Octahedra})} \end{array}$ & $\begin{array}{c} {96}\\ {({\rm  Triangles})} \end{array}$ & $\begin{array}{c} {96}\\ {} \end{array}$ & $\begin{array}{c} {24}\\ {} \end{array}$ & $\begin{array}{c} {{\rm Cube }}\\ {} \end{array}$ & $\begin{array}{c} {\{343\} }\\ {} \end{array}$  \\
$\begin{array}{c}  {120{\rm -cell}}\\ {({\rm Dodecahedra})} \end{array}$ & $\begin{array}{c} {720}\\ {({\rm Pentagons})}  \end{array}$ & $\begin{array}{c} {1200}\\ {} \end{array}$ & $\begin{array}{c} {600}\\ {} \end{array}$ & $\begin{array}{c} {{\rm Tetrahedron }}\\ {} \end{array}$ & $\begin{array}{c} {\{533\} }\\ {} \end{array}$ \\
$\begin{array}{c} {600{\rm -cell}}\\ {({\rm Tetrahedra})}  \end{array}$ & $\begin{array}{c} {1200}\\ {({\rm Triangles})}  \end{array}$ & $\begin{array}{c} {720}\\ {} \end{array}$ & $\begin{array}{c} {120}\\ {}  \end{array}$ & $\begin{array}{c} {{\rm Icosahedron }}\\ {} \end{array}$ & $\begin{array}{c} {\{335\} } \\{} \end{array}$ \\
\hline
\end{tabular}
\end{center}
\caption{
\textit{All possible lattices that can be constructed on a 3-sphere using the regular convex polychora.
The Schl\"{a}fli symbols $\{pqr\}$ denote the number of edges to a face, $p$, the number of faces 
that meet at the vertex of a cell, $q$, and the number of cells that meet at an edge, $r$.}}
\label{table1}
\end{table*}

\subsection{The 5-Cell} 

Let us first consider the 5-cell.  This is the smallest lattice in Table \ref{table1}, and has masses positioned as specified in Table \ref{table2}. The source functions, $f_i$, from Eq. (\ref{init2}), are then given by $f_i = \sin \left[ \textstyle\frac{1}{2} \cos^{-1} (h_i)\right]$, where the functions $h_i$ are given by\footnote{This corrects a typo in $h_3$ in the published version of \cite{lattice1}.}
\ba
h_1&=&\cos\chi\\
h_2&=&\frac{\sqrt{15}}{4} \cos \theta \sin \chi -\frac{\cos \chi}{4}\\
 h_3&=&\sqrt{\frac{5}{6}} \sin \chi \sin \theta \cos \phi \nonumber\\&&-\sqrt{\frac{5}{48}} \sin \chi \cos \theta -\frac{\cos \chi}{4}\\
h_4&=&\sqrt{\frac{5}{6}} \sin \chi \sin \theta \sin \left(\phi-\frac{\pi}{6} \right)\nonumber\\&&-\sqrt{\frac{5}{48}} \sin  \chi \cos \theta -\frac{\cos \chi}{4}\\
h_5&=&-\sqrt{\frac{5}{6}} \sin \chi \sin \theta \sin \left(\phi+\frac{\pi}{6} \right)\nonumber\\&&-\sqrt{\frac{5}{48}} \sin \chi \cos \theta -\frac{\cos \chi}{4} .
\ea

\begin{table}[h!]
\begin{center}
\begin{tabular}{|c|l|}
\hline
{\bf Point} & {\bf ($\chi$,
  $\theta$, $\phi$)}\\
\hline
$(i)$ & $\left(0, \frac{\pi}{2}, \frac{\pi}{2}
\right)$ \\
$(ii)$ & 
$\left(\cos^{-1}(-\frac14), 0, \frac{\pi}{2}
\right)$ \\
$(iii)$   &  $\left(\cos^{-1}(-\frac14),
\cos^{-1}(-\frac13), 0
\right)$ \\
$(iv)$ 
& $\left(\cos^{-1}(-\frac14),\cos^{-1}(-\frac13) , \frac{2 \pi}{3}\right)$ \\
$(v)$ 
& $\left(\cos^{-1}(-\frac14),\cos^{-1}(-\frac13) , \frac{4 \pi}{3} \right)$ \\
\hline
\end{tabular}
\end{center}
\caption{{\protect{\textit{Coordinates {\rm ($\chi,\theta,\phi$)} of the positions of the masses on the background 3-sphere. In this table, and
throughout, $\cos^{-1}$ refers to the inverse cosine, and not its reciprocal.}}}}
\label{table2}
\end{table}

\subsection{The 8-Cell}

Let us now consider the 8-cell. The positions of the masses in this case are given in Table \ref{table3}. The $f_i$ from Eq. (\ref{init2}) are then
found to be 
\ba
f_1 &=& \sin \left[\frac{\chi}{2}\right]\\
f_2 &=& \cos \left[\frac{\chi}{2}\right]\\
f_3 &=& \sin \left[ \frac{1}{2} \cos^{-1} \left( \cos \theta
  \sin \chi \right) \right]\\
f_4 &=& \cos \left[ \frac{1}{2} \cos^{-1} \left( \cos \theta
  \sin \chi \right) \right]\\
f_5 &=& \sin \left[ \frac{1}{2} \cos^{-1} \left( \cos \phi \sin \theta 
  \sin \chi \right) \right]\\
f_6 &=& \cos \left[ \frac{1}{2} \cos^{-1} \left( \cos \phi \sin \theta
  \sin \chi \right) \right]\\
f_7 &=& \sin \left[ \frac{1}{2} \cos^{-1} \left( \sin \phi \sin \theta
  \sin \chi \right) \right]\\
f_8 &=& \cos \left[ \frac{1}{2} \cos^{-1} \left( \sin \phi \sin \theta 
  \sin \chi \right) \right].
\ea

\begin{table}[h!]
\begin{center}
\begin{tabular}{|c|l|}
\hline
\bf{Point}  & \bf{($\chi$,
  $\theta$, $\phi$)}\\ 
\hline
$(i)$ &  $\left(0, \frac{\pi}{2}, \frac{\pi}{2}
\right)$ \\
$(ii)$ &  $\left(\pi, \frac{\pi}{2}, \frac{\pi}{2}
\right)$ \\
$(iii)$ &  $\left( \frac{\pi}{2}, 0, \frac{\pi}{2}
\right)$ \\
$(iv)$   &  $\left( \frac{\pi}{2}, \pi, \frac{\pi}{2}
\right)$ \\
$(v)$  &  $\left( \frac{\pi}{2}, \frac{\pi}{2}, 0
\right)$ \\
$(vi)$ &  $\left( \frac{\pi}{2}, \frac{\pi}{2}, \pi
\right)$ \\
$(vii)$ &  $\left( \frac{\pi}{2}, \frac{\pi}{2}, \frac{\pi}{2}
\right)$ \\
$(viii)$  &  $\left( \frac{\pi}{2}, \frac{\pi}{2},
\frac{3 \pi}{2} \right)$ \\
\hline
\end{tabular}
\end{center}
\caption{{\protect{\textit{Coordinates  {\rm
($\chi,\theta,\phi$)} of the positions of the masses on the background 3-sphere.}}}}
\label{table3}
\end{table}

\subsection{Models with 16-600 Equally Spaced Masses}

The coordinate positions and functions associated with the 16, 24, 120 and 600 equally spaced masses in the other lattices will not be presented explicitly here. They are constructed in a similar way to the cases discussed in detail above. Visual representations of all of these structures are presented in reference \cite{lattice1}, and the evolution of the initial data found there will be studied in what follows.

\section{Curves With Local Rotational Symmetry}
\label{sec:lrs}

There are a number of curves within each of the lattices described in Sec. \ref{sec:lattices} about which there exists an $n$-fold symmetry under a set of discrete rotations. What we mean by this is that there exist curves along which there exists a space-like direction (the axis of symmetry) that can be aligned with one of our orthonormal basis vectors, such that all covariantly defined quantities that are picked out by the geometry, and that are written in terms of an orthonormal tetrad that contains this vector, are invariant under spatial rotations that leave this vector unchanged.  That is, there exists a space-like unit vector ${\bf e}_1$ such that under the transformation ${\bf \tilde{e}}_1={\bf e}_1$, ${\bf \tilde{e}}_0={\bf e}_0$ and
\ba
\label{rot1}
{\bf \tilde{e}}_2 &=& {\bf e}_2 \cos \phi_m - {\bf e}_3 \sin \phi_m\\
{\bf \tilde{e}}_3 &=& {\bf e}_3 \cos \phi_m + {\bf e}_2 \sin \phi_m
\label{rot2}
\ea
all covariantly defined quantities picked out by the geometry remain identical before and after the transformation:
\be
\label{disrot}
T^{\tilde{a} \tilde{b} \tilde{c} \dots}_{\phantom{\tilde{a} \tilde{b} \tilde{c} \dots} \tilde{d} \tilde{e} \tilde{f} \dots} = T^{a  b c \dots}_{\phantom{a b c \dots} d e f \dots}.
\ee
where $T$ is a tensor, and where indices with a tilde denote those corresponding to the set of basis vectors after the rotation. Here $\phi_m$ denotes a set of angles that lie in the interval $(0,2 \pi)$.

\begin{figure}[t!]
\begin{centering}
\includegraphics[width=3.5in]{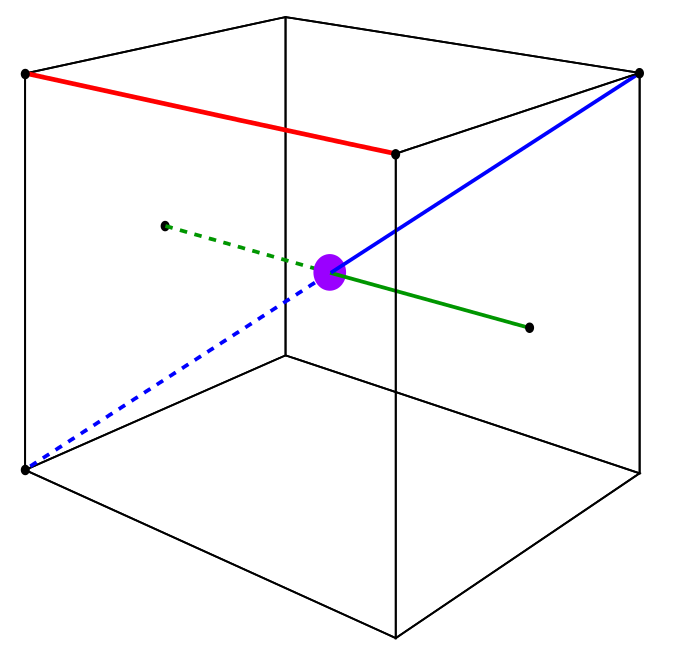}
\par\end{centering}
\caption{An illustration of a cubic lattice cell. An example edge is highlighted in red, an example curve that connects cell centres through the centre of a cell face is highlighted in green, and an example curve that connects cell centres through a vertex is highlighted in blue. Each of these curves exhibits manifest discrete rotational symmetries, due to the symmetries of the cell, and the identical nature of all other cells.}
\centering{}\label{cubefig}
\end{figure}

Examples of curves that display this type of symmetry, and that we will make use of in the following sections, are the edges of cells, the curves that connect cell centres through cell faces, and the curves that connect cell centres with vertices. These curves are highlighted in Fig. \ref{cubefig} using the example of a cubic lattice cell. In each of these cases the angles $\phi_m$ are given by
\be
\phi_{m} = \frac{2 \pi m}{n},
\ee
where $m=1, \dots ,n-1$, and $n$ is an integer. For the edges of the cells in the six lattices discussed in Sec. \ref{sec:lattices} we have $n=3$, $3$, $4$, $3$, $3$ and $5$ for the $5$-cell, $8$-cell, $16$-cell, $24$-cell, $120$-cell and $600$-cell, respectively. For the curves that connect cell centres, and that pass through the centre of cell faces, we have $n=3$, $4$, $3$, $3$, $5$, and $3$, while for the curves that connect cell centres with vertices we have $n=3$, $3$, $3$, $4$, $3$ and $3$. These numbers can all be read off from the Schl\"{a}fli symbols in Table \ref{table1}.

We will now show that these discrete symmetries, which must exist because of the regularity of our lattices, imply that there also exists a continuous rotational symmetry about the same set of curves. To do this, let us first consider a vector at a point on one of these curves, $T_{\alpha}$. From Eq. (\ref{disrot}) we then have that $T_2 = T_{\tilde{2}}$ and $T_3 = T_{\tilde{3}}$, or, using Eqs. (\ref{rot1}) and (\ref{rot2}), that
\ba
(1- \cos \phi_{m}) T_2-\sin \phi_{m} T_3 &=& 0 \nonumber\\
\sin \phi_{m} T_2- (1- \cos \phi_{m}) T_3 &=& 0.\nonumber
\ea
This is in fact a set of $2 (n-1)$ equations, which for $n \geq 2$ admits as the only solution
\be
\label{t20}
T_2 = T_3 = 0.
\ee
Therefore no space-like directions orthogonal to ${\bf e}_1$ can be picked out by any covariantly defined vector, and we say that the spacetime is Locally Rotationally Symmetric (LRS) about ${\bf e}_1$ along the curve we are considering. This should not be taken to imply that the spacetime is LRS about every point, but rather that there exists a network of LRS curves that permeate the spacetime.

The consequences of symmetry under discrete rotations can also be demonstrated for symmetric rank-2 tensors, $T_{\alpha \beta}$. We see that all components that contain a single index that is $2$ or $3$ must vanish, for the same reason as the vector components above. The remaining components are then given by $T_{11}$, $T_{22}$, $T_{33}$ and $T_{23}$. The requirements that $T_{\tilde{2} \tilde{2}}=T_{22}$, $T_{\tilde{3} \tilde{3}}=T_{33}$ and $T_{\tilde{2} \tilde{3}} = T_{23}$ then give
\ba
\sin^2 \phi_m T_{22} &=& \sin^2 \phi_{m} T_{33} - 2 \sin \phi_{m} \cos \phi_{m} T_{23}  \nonumber\\ \nonumber
\sin^2 \phi_m T_{33} &=& \sin^2 \phi_{m} T_{22} + 2 \sin \phi_{m} \cos \phi_{m} T_{23}  \nonumber\\ \nonumber
\sin \phi_m \cos \phi_{m} T_{33} &=& \sin \phi_{m} \cos \phi_{m} T_{22}  - 2 \sin^2 \phi_{m} T_{23}. \nonumber
\ea
For $n \geq 3$ these equations give $T_{23}=0$ and $T_{22}=T_{33}$. This again shows that no space-like directions orthogonal to ${\bf e}_1$ can be picked out, as expected for a point displaying rotational symmetry around ${\bf e}_1$. For $n=2$ we can appeal to Eq. (\ref{t20}), which implies that rank-2 tensors have no fixed eigenvector direction. This means that eigenvectors must be degenerate, and therefore that $(T_{22}-T_{33})^2 + 4 (T_{23})^2=0$, which gives the same conditions as stated above.

These properties mean that the only non-zero {kinematic and geometric} quantities in irrotational, geodesic vacuum spacetimes at LRS points are \cite{Ellis-67, Stewart-Ellis-68, vanElst-Uggla-97}
\be
\nonumber
\Theta, \; \sigma_{+}, \; E_{+} \quad {\rm and} \quad H_{+}.
\ee
{Moreover}, the LRS symmetry implies (i)  that ${\bf e}_2 \cdot \nabla_1 {\bf e}_1 = {\bf e}_3 \cdot \nabla_1 {\bf e}_1 =0$, as $\nabla_1 {\bf e}_1$ is a covariantly defined vector picked out by the geometry, and the $2$ and $3$ components of any such vector are already known to vanish, (ii) that the integral lines of ${\bf e}_1$ must be shear-free, as any symmetric rank-2 tensor must have vanishing off-diagonal components and $ T_{22}=T_{33}$, and (iii) that all covariantly defined scalars must have vanishing frame derivatives in the $2$ and $3$ directions, as these derivatives are themselves vectors. These results imply that (i) $\gamma^1_{\phantom{1}21} =0 =\gamma^1_{\phantom{1}31} $, and (ii) $\gamma^2_{\phantom{2}12} = \gamma^3_{\phantom{3}13}$ and $\gamma^3_{\phantom{3}12} = \gamma^2_{\phantom{2}13}$, which in turn imply that (i) $a_2 -n_{13} = 0 =a_3 +n_{12}$, and that (ii) $n_{23}=0$ and $n_{22} + n_{33}=0$.

As well as rotational symmetry, the cell faces that meet at the curves we have been considering must also admit a reflection symmetry. This additional symmetry means that skew symmetric rank-2 tensors, $S_{\alpha \beta}$, must vanish on these curves. This can be seen by recognising that $S_{1 2}$ = $S_{1 3}$=0, from the arguments above, and that by choosing ${\bf e}_2$ to lie in the symmetry plane we must have that all quantities remain invariant under ${\bf e}_3 \rightarrow -{\bf e}_3$. This gives
\be
S_{23} = -S_{23} = 0,
\ee
and so $S_{\alpha \beta}=0$ for all $\alpha$ and $\beta$. In particular, this means (i) that the integral lines of ${\bf e}_1$ must be irrotational, as vorticity is a skew symmetric tensor, and (ii) that ${\bf e}_1$ contracted with the curl of any covariantly defined vector must vanish, where curl is defined here as
\be
({\rm curl} \; V)^{\alpha} \equiv \epsilon^{\alpha \beta \gamma} {\bf e}_{\beta} (V_{\gamma}).
\ee
The second of these results follows from the first as the vanishing of the vorticity of ${\bf e}_1$ means that ${\bf e}_{[2}  (V_{3]}) = e_{[2}^{\phantom{[2} \nu} e_{3]}^{\phantom{3]} \mu} V_{\mu ; \nu}= e_{2}^{\phantom{2} \nu} e_{3}^{\phantom{3} \mu} V_{[\mu ; \nu]}$. These results mean that (i) $\gamma^1_{\phantom{1} 23} =0$, and hence $n_{11}=0$, and that (ii) ${\bf e}_{[2} \left( \sigma_{3]1}\right)=0$, which will be of use below.

\subsection{Evolution Equations of LRS Curves in Orthonormal Frame Variables}

Under the restrictions described above it can be seen that Eq. (\ref{con3}) gives
\be
\label{Hlrs}
H_{+} = 0,
\ee
which results in Eqns. (\ref{evo1}), (\ref{evo2}) and (\ref{evo4}) giving
\ba
\label{Tlrs}
{\bf e}_0 (\Theta ) &=& - \frac{1}{3}\Theta^2 - \frac{2}{3} (\sigma_{+})^2\\
\label{Slrs}
{\bf e}_0 (\sigma_{+}) &=& - \frac{1}{3} (2 \Theta - \sigma_{+}) \sigma_{+} - E_{+}\\
\label{Elrs}
{\bf e}_0 (E_{+}) &=& -(\Theta +\sigma_{+}) E_{+},
\ea
where we have used ${\bf e}_{[2} (H_{3]1})  = 0$ in Eq. (\ref{Elrs}).
By defining two new variables equal to the Hubble expansion rates in the directions parallel and perpendicular to the LRS curve, via
\ba
\mathcal{H}_{||} &\equiv& \theta_{11} = \frac{1}{3} \Theta +\sigma_{11} = \frac{1}{3} (\Theta -2 \sigma_{+})\\
\mathcal{H}_{\perp} &\equiv& \theta_{22} = \frac{1}{3} \Theta +\sigma_{22} = \frac{1}{3} (\Theta + \sigma_{+}),
\ea
we find that these equations can be decoupled into
\ba
\label{Hp}
{\bf e}_0 (\mathcal{H}_{\perp}) + \mathcal{H}_{\perp}^2 &=& - \frac{1}{3} E_{+}\\
\label{Ep}
{\bf e}_0 (E_{+}) + 3 \mathcal{H}_{\perp} E_{+} &=& 0 ,
\ea
and
\be
\label{H11}
{\bf e}_0 (\mathcal{H}_{||}) + \mathcal{H}_{||}^2 = \frac{2}{3} E_{+}.
\ee
Eqs. (\ref{Hp}) and (\ref{Ep}) are closely analogous to the Friedmann and energy conservation equations of FLRW cosmology. This analogy is made all the stronger by the fact that $\mathcal{H}_{\perp}$ and $\mathcal{H}_{||}$ are the rates of expansion in the directions perpendicular to, and parallel with, the curves that exhibits local rotational symmetry (see e.g. \cite{Ellis-vanElst-1998}). We will explain this further below.

\subsection{Evolution of a LRS Curve}

Let us consider the lattice models constructed in Section \ref{sec:lattices}. If we rotate the lattice until the curve that interests us has $\theta=$constant and $\phi=$constant, then the length of that curve is given by
\be
\label{ell}
\ell (t) = \int_{\chi_1}^{\chi_2} \sqrt{g_{\chi\chi}} d \chi,
\ee
where $\chi_1$ and $\chi_2$ denote the values of the $\chi$ coordinate at its two ends.

To determine the evolution of $\ell (t)$ requires knowledge of the evolution of $g_{\chi \chi}$ at every point along the curve, which is given by $\theta_{\chi \chi} = \frac{1}{2} {\bf e}_0 ( g_{\chi \chi})$. The relationship between $\theta_{\chi \chi}$ and the corresponding expression in orthonormal frame components is $\theta_{\chi \chi}= g_{\chi \chi} \theta_{11}$. It then follows that $\theta_{11} = \frac{1}{2} {\bf e}_0 ( \ln g_{\chi \chi})$, which can be integrated to give $g_{\chi \chi} \propto a_{||}^2$, where $a_{||}$ is defined implicitly by
\be
\label{a11}
\theta_{11} \equiv  {\bf e}_0 ( \ln a_{||}) =   \frac{\dot{a}_{||}}{a_{||}},
\ee
and where in the last expression we use an over-dot to denote a derivative in the direction ${\bf e}_0$. The evolution of $\ell (t)$ is then given by
\be
\label{edge}
\ell (t) = \int_{\chi_1}^{\chi_2} a_{||} \sqrt{(g_{\chi\chi})_0} \;  d\chi,
\ee
where $ (g_{\chi\chi})_0$ is the value of $g_{\chi \chi}$ at the moment when $a_{||}=1$. The variable $a_{||}$ defined in Eq. (\ref{a11}) is therefore the scale factor along the LRS curve.

The evolution of $a_{||}$ is given by Eq. (\ref{H11}), which in the notation used in this section is given by
\be
\label{H112}
\frac{\ddot{a}_{||}}{a_{||}} = \frac{2}{3} E_{+},
\ee
where over-dots denote partial derivatives with respect to $t$ (the proper time along integral curves of $u^{\mu}$). Eqs. (\ref{Hp}) and (\ref{Ep}) can similarly be written as 
\be
\label{Hp2}
\frac{\ddot{a}_{\perp}}{a_{\perp}} = -\frac{1}{3} E_{+}
\ee
and
\be
\label{Ep2}
\dot{E}_{+} + 3 \frac{\dot{a}_{\perp}}{a_{\perp}} E_{+} =0,
\ee
where $a_{\perp}$ is defined implicitly by $\theta_{22} \equiv {\bf e}_0(\ln a_{\perp})$. Equation (\ref{Ep2}) can be integrated to give $E_{+}=(E_{+})_0 /a_{\perp}^3$, where the subscript $0$ again denotes that the quantity in brackets is to be evaluated at the moment when $a_{\perp}=1$. This shows that $E_{+}$ evolves as a function of $a_{\perp}$ in the same way that the energy density of a pressure-less fluid evolves with respect to the scale factor in FLRW cosmology.

Integrating Eq. (\ref{Hp2}) then gives
\be
\label{F1}
\frac{\dot{a}_{\perp}^2}{a_{\perp}^2} = \frac{\frac{2}{3} (E_{+})_0}{a_{\perp}^3} - \frac{k}{a_{\perp}^2},
\ee
where $k$ is a constant of integration. If we now exploit the invariance under rescaling of $a_{\perp}$ in Eqs. (\ref{Hp2}) and (\ref{Ep2}) then we can arrange that $\dot{a}_{\perp}=0$ when $a_{\perp}=1$. Quantities with subscript $0$ now correspond to their value on the time-symmetric hypersurface that we specified our initial data on, as we must have $\dot{a}_{\perp}=0$ at this time (because $\Theta=\sigma_{+}=0$, and so $\theta_{22}=0$). In this case we have from Eq. (\ref{F1}) that $k=\frac{2}{3} (E_{+})_0$. In what follows we will assume that the scale invariance of $a_{\perp}$ has been used to assign this value to $k$. The form of the solutions to Eq. (\ref{F1}) then depends only on the sign and magnitude of $(E_{+})_0$.

\subsubsection{Evolution when ${ (E_{+})_0=0}$}

{When $(E_{+})_0=0$, Eqs.} (\ref{H112}) and (\ref{F1}) give
\ba
a_{||} &=& 1\\
a_{\perp} &=&1,
\ea
where we have used the fact that we must also have $\dot{a}_{||}=0$ when $a_{\perp}=1$, and that the invariance under rescaling of $a_{||}$ in Eq. (\ref{H112}) can also be used to set $a_{||}=1$ at this time. The LRS curves do not therefore evolve at points where ${(E_{+})_0=0}$, which in fact corresponds to the condition that the geometry of spacetime at any such points must be Riemann flat, and stay Riemann flat forever. These regions are therefore strongly locally isometric with Minkowski space for all time.

\subsubsection{Evolution when ${ (E_{+})_0>0}$}

\begin{figure}[t!]
\begin{centering}
\includegraphics[width=4in]{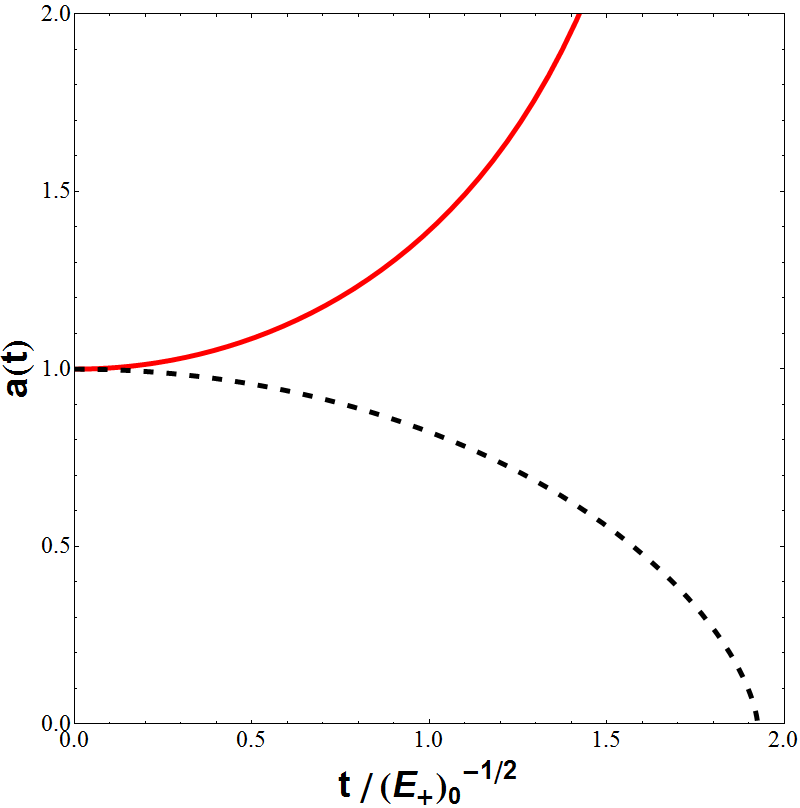}
\par\end{centering}
\caption{The scale factors $a_{\perp}(t)$ (dotted black line) and $a_{||}(t)$ (solid red line), when $(E_{+})_0>0$.}
\centering{}\label{a+fig}
\end{figure}

When $(E_{+})_0>0$ the solution to Eq. (\ref{F1}) can be written in parametric form as
\ba
\label{aperp+}
a_{\perp} &=& \cos^2 \eta\\
\label{t+}
t-t_0 &=& \frac{1}{\sqrt{\frac{2}{3} (E_{+})_0}} \left( \eta +\frac{1}{2} \sin (2\eta) \right),
\ea
where $-\frac{\pi}{2} < \eta < \frac{\pi}{2}$, and where $t_0$ is constant of integration, defined such that $t=t_0$ on the time-symmetric hypersurface. We can now solve Eq. (\ref{H112}) to give the scale factor along the LRS curves as
\be
\label{a||+}
a_{||} = \frac{1}{2} \left( 3 -\cos^2 \eta +3 \eta \tan \eta \right),
\ee
where we have used $\dot{a}_{||}=0$ on the initial hypersurface, and where we have also used the scale invariance to set $a_{||}=1$ at that time. The functions in Eqs. (\ref{aperp+})-(\ref{a||+}) are shown graphically in Fig. \ref{a+fig}.

\subsubsection{Evolution when ${ (E_{+})_0<0}$}

\begin{figure}[t!]
\begin{centering}
\includegraphics[width=4in]{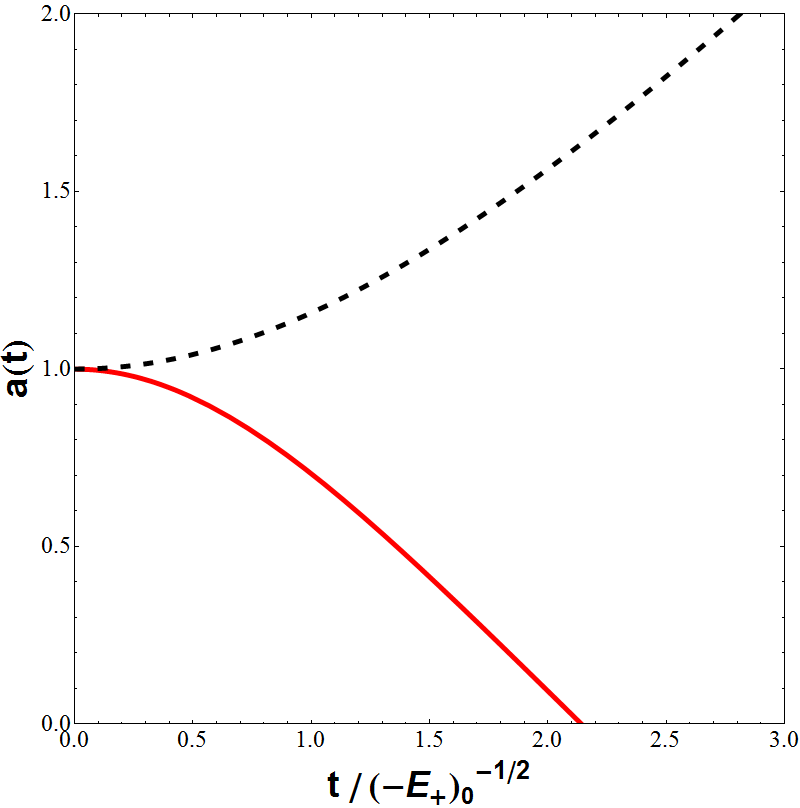}
\par\end{centering}
\caption{The scale factors $a_{\perp}(t)$ (dotted black line) and $a_{||}(t)$ (solid red line), when $(E_{+})_0<0$.}
\centering{}\label{a-fig}
\end{figure}

Similarly, when $(E_{+})_0<0$ the solutions to Eqs. (\ref{F1}) and (\ref{H112}) can be written as
\ba
\label{aperp-}
a_{\perp} &=& \cosh^2 \eta\\
\label{a||-}
a_{||} &=& \frac{1}{2} \left( 3 -\cosh^2 \eta -3 \eta \tanh \eta \right)
\ea
where
\be
\label{t-}
t-t_0 = \frac{1}{\sqrt{-\frac{2}{3} (E_{+})_0}} \left( \eta +\frac{1}{2} \sinh (2\eta) \right),
\ee
and where $t_0$ is again a constant of integration that has been defined such that $t=t_0$ on the time-symmetric hypersurface. The functions in Eqs. (\ref{aperp-})-(\ref{t-}) are shown graphically in Fig. \ref{a-fig}.

\section{Evolution of the Edge Lengths}
\label{sec:edges}

\subsection{Evolution of ${\bf E_{+}}$}

\begin{figure}[t!]
\begin{centering}
\includegraphics[width=4.1in]{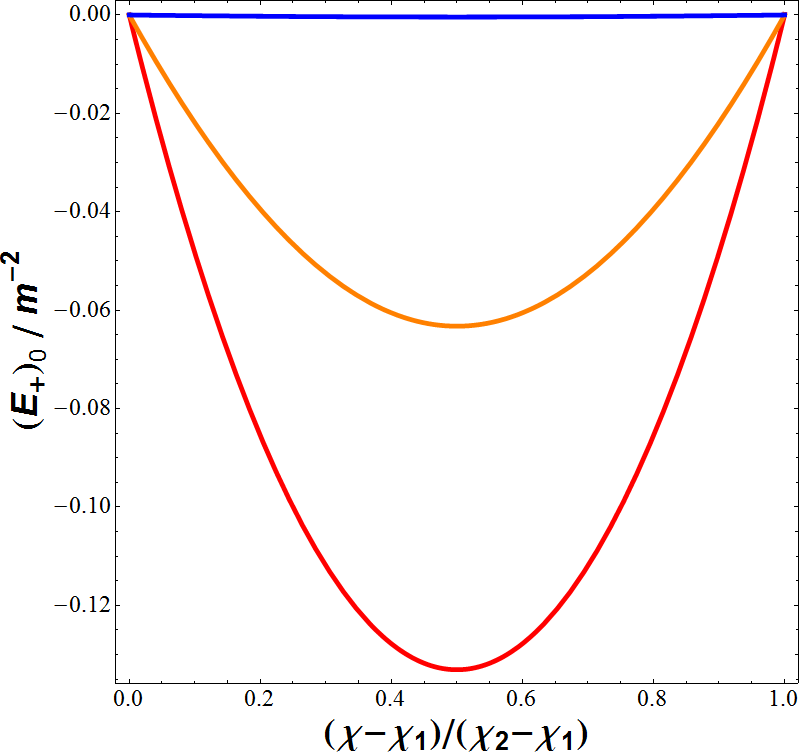}
\par\end{centering}
\caption{The value of $(E_{+})_0$ along an edge, in units of $m^{-2}$. The lower curve (red) is for the 5-cell, the middle curve (orange) is for the 8-cell, and the upper curve (blue) is for the 120-cell.}
\centering{}\label{E1fig}
\end{figure}

\begin{figure}[t!]
\begin{centering}
\includegraphics[width=4.1in]{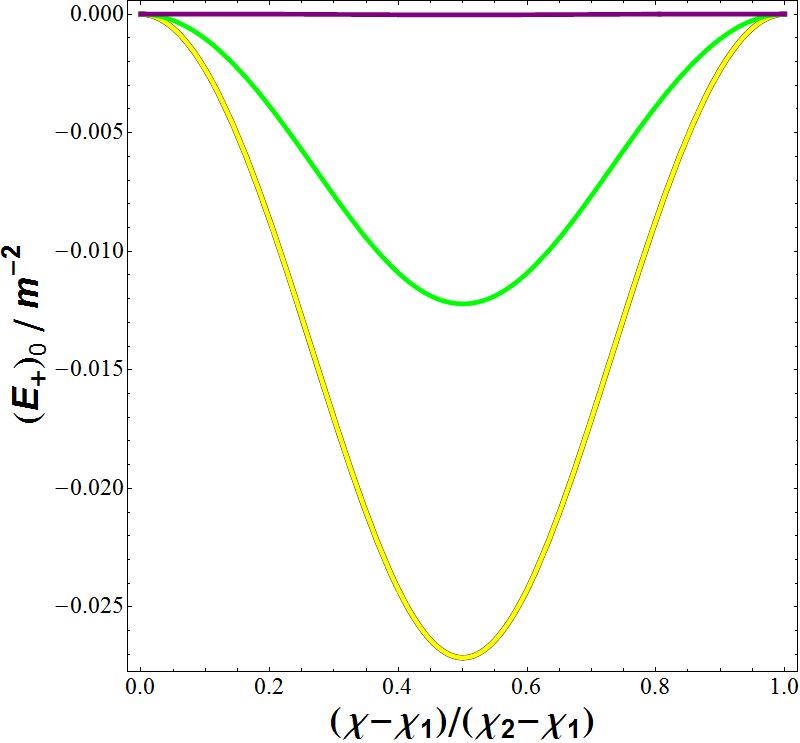}
\par\end{centering}
\caption{The same quantity shown in Fig. \ref{E1fig}, but now the lower curve (yellow) is for the 16-cell, the middle curve (green) is for the 24-cell, and the upper curve (purple) is for the 600-cell.}
\centering{}\label{E2fig}
\end{figure}

The first step in performing the integral in Eq. (\ref{edge}) is to calculate $(E_{+})_0$ as a function of position along the edge. We will use the coordinate $\chi$ to denote positions along the edge, and will obtain an expression for $(E_{+})_0$ as a function of $\chi$ by using Eqs. (\ref{Eab}), (\ref{init1}) and (\ref{init2}). The profiles for the six lattices that we introduced in Section \ref{sec:lattices} are shown graphically in Figs. \ref{E1fig} and \ref{E2fig}. Here we have taken each mass, in each of the lattices, to take an identical value, $m$. The quantities $\chi_1$ and $\chi_2$ in these plots correspond to the coordinate positions of the vertices at the ends of the edges, as in Eq. (\ref{edge}).

It can be seen from Figs. \ref{E1fig} and \ref{E2fig} that as the number of masses is increased, the amplitude of $(E_{+})_0/m^{-2}$ decreases. It can also be seen that there are two qualitatively different profiles possible (which is the reason we have presented the profiles of these six lattices on two plots, rather than one). In Fig. \ref{E1fig} we have displayed the results for the 
{5, 8 and the 120-cell models}. In each of these cases we can see that while $(E_{+})_0=0$ at 
{the ends of the edges, the gradient of this quantity is non-zero at these points}. 
In contrast, the results for the {16, 24 and 600-cell models} that are shown in Fig. \ref{E2fig} have both $(E_{+})_0=0$ and a zero first derivative of this quantity at the ends of each edge. 

The reason for this different behaviour can be deduced from the vertex figures listed in Table 1, and is due to the way edges meet at the vertices in the different lattices. If two edges are contiguous (i.e. meet end-to-end, so that they form one continuous smooth curve), then the profile of $(E_{+})_0$ along each edge must have zero first derivative at the ends of the edge in order for the geometry to be smooth, and for there to be mirror symmetry about the vertex. This is the case for the lattices in Fig. \ref{E2fig}. If the edges are not contiguous (i.e. if the curve that is the extension of an edge does not overlap with the edge of a neighbouring cell) then the first derivative does not have to vanish, as the value of $(E_{+})_0$ can change sign as one passes through the vertex. This is the case for the lattices in Fig. \ref{E1fig}. Later on, we will find that these two different types of profiles have significant consequences for the evolution of the edge lengths as a whole.

\begin{figure}
\centering
  \subfloat[$E_{+}$ along an edge in the 5-cell.]{\label{heat5fig}
    \includegraphics[width=4.1in]{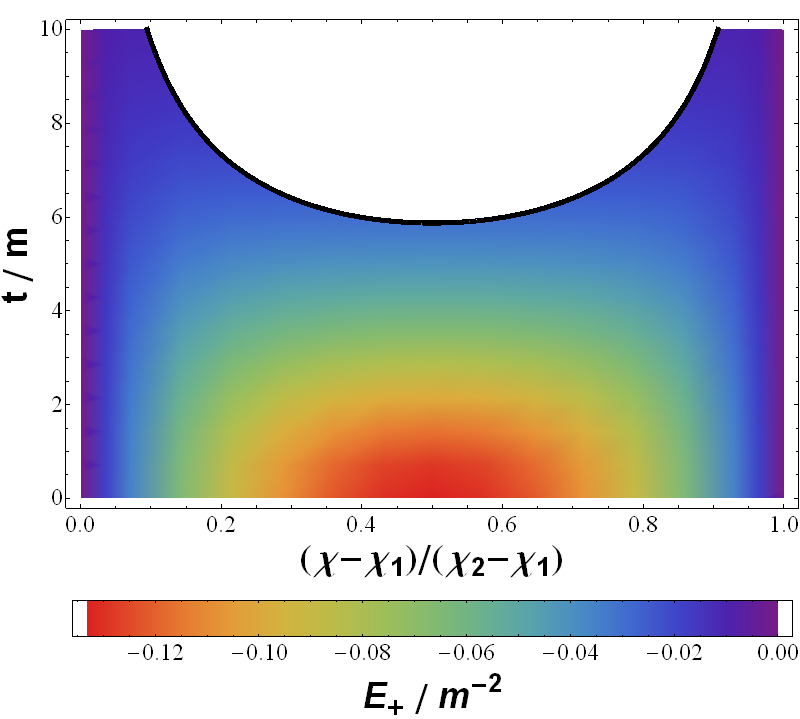}}\qquad\qquad
  \subfloat[$E_{+}$ along an edge in the 8-cell.]{\label{heat8fig}
    \includegraphics[width=4.1in]{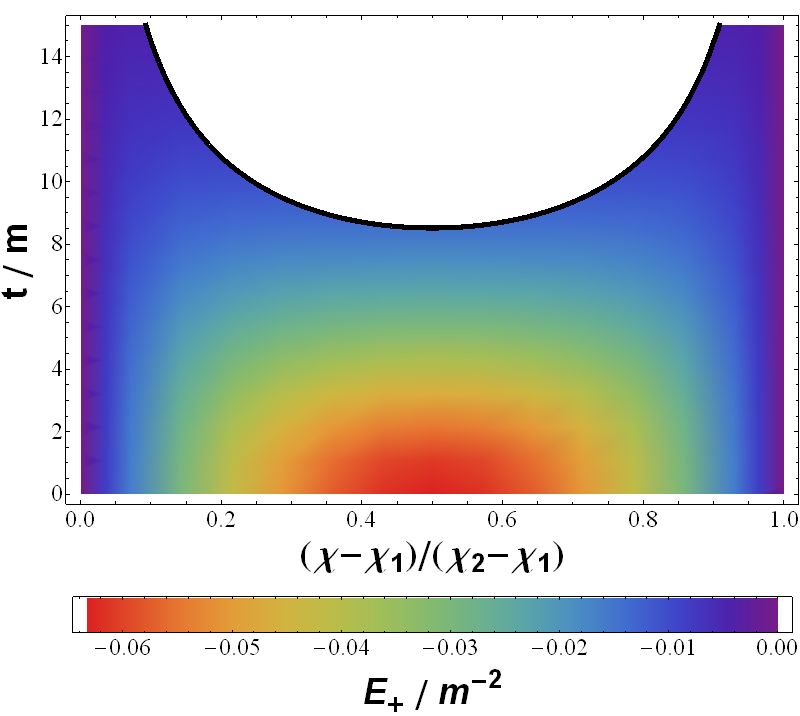}}
  \caption{Spacetime diagrams of the value of $E_{+}$ along the edge of each of our lattice cells. Excluded regions, above the black line, have passed the point at which $a_{||}=0$.}
\end{figure}

\begin{figure}
\centering
  \subfloat[$E_{+}$ along an edge in the 16-cell.]{\label{heat16fig}
    \includegraphics[width=4.1in]{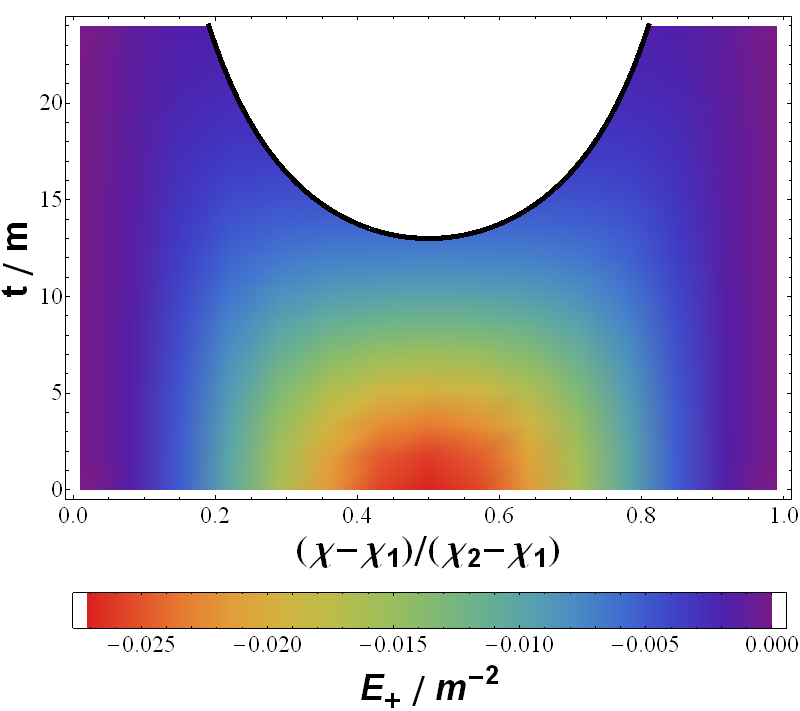}}\qquad\qquad
  \subfloat[$E_{+}$ along an edge in the 24-cell.]{\label{heat24fig}
  \includegraphics[width=4.1in]{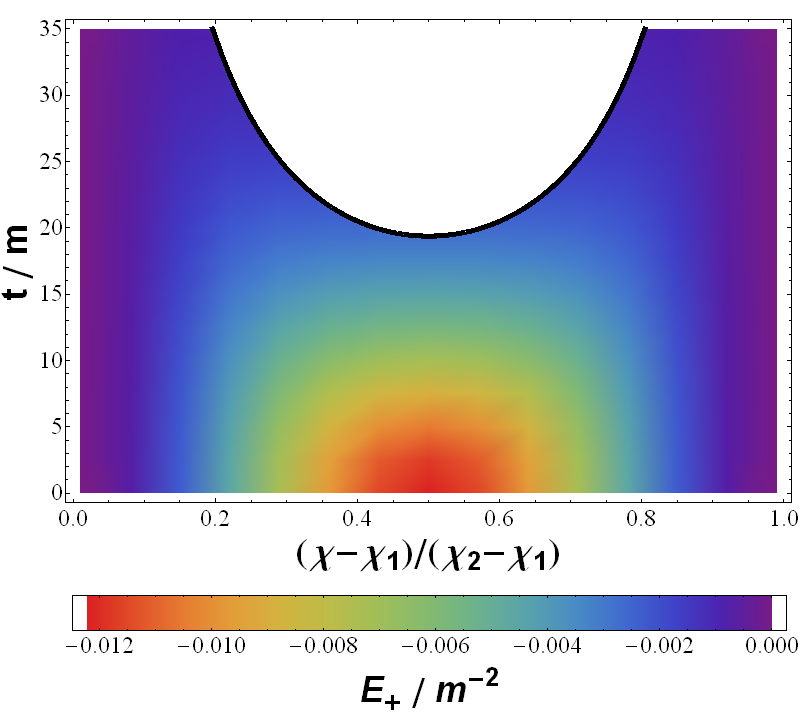}}
 \caption{Spacetime diagrams of the value of $E_{+}$ along the edge of each of our lattice cells. Excluded regions, above the black line, have passed the point at which $a_{||}=0$.}
\end{figure}

\begin{figure}
\centering
  \subfloat[$E_{+}$ along an edge in the 120-cell.]{\label{heat120fig}
    \includegraphics[width=4.1in]{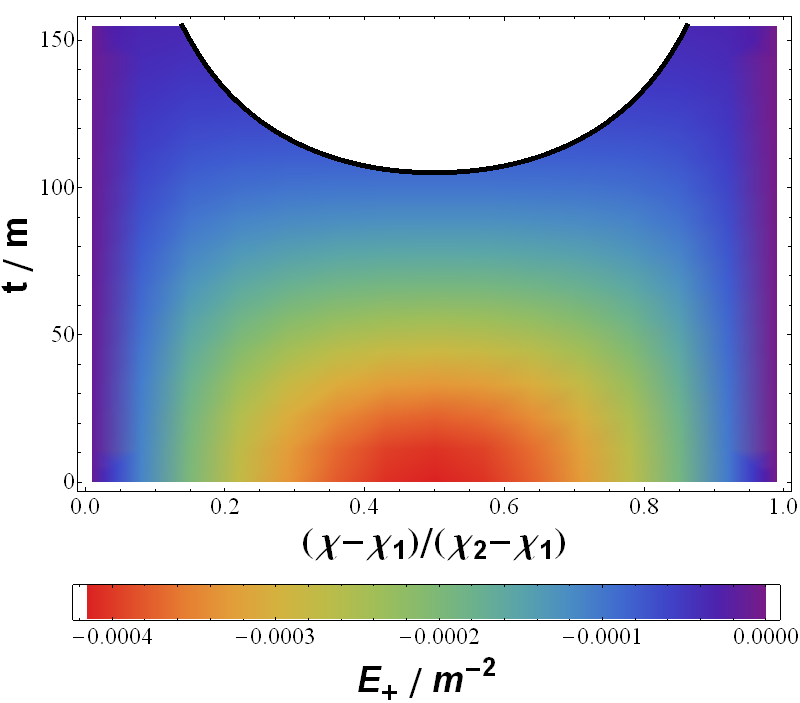}}\qquad\qquad
  \subfloat[$E_{+}$ along an edge in the 600-cell.]{\label{heat600fig}
  \includegraphics[width=4.1in]{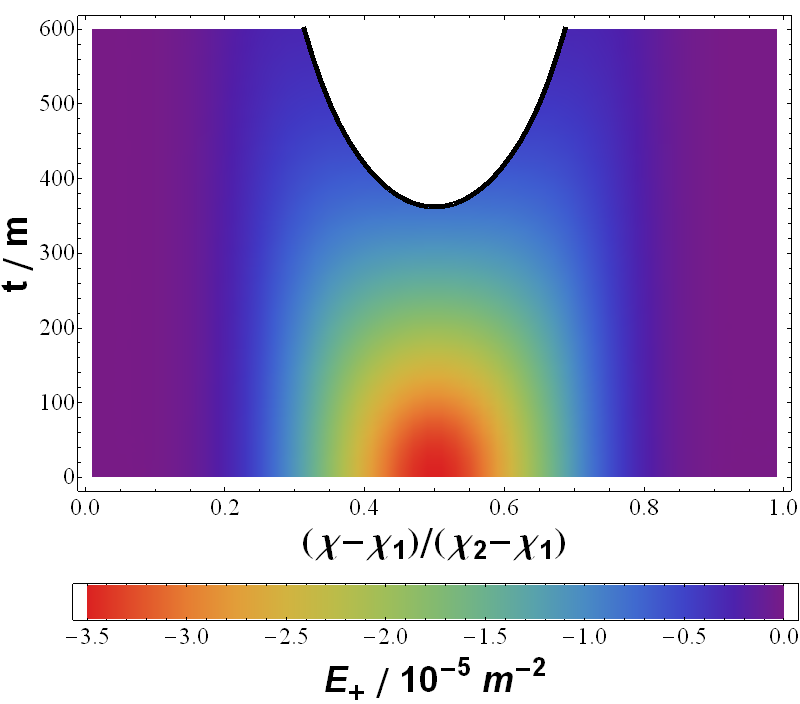}}
  \caption{Spacetime diagrams of the value of $E_{+}$ along the edge of each of our lattice cells. Excluded regions, above the black line, have passed the point at which $a_{||}=0$.}
\end{figure}

In Figs. \ref{heat5fig}-\ref{heat600fig} we plot the evolution of $E_{+}$ along the full length of an edge, in each of our lattices. In these plots the regions in which $a_{||} < 0$ have been excluded. For $E_{+} \neq 0$ it is the case that $a_{||} =0$ occurs at finite time. This does not, however, correspond to a curvature singularity, as it can be seen from Eqs. (\ref{Hp2}) and (\ref{Ep2}) that when $E_{+}<0$ (as is the case at all points along any edge) we have $\ddot{a}_{\perp} >0$. This means that $a_{\perp}$ is always expanding as one evolves away from the time-symmetric initial surface. If $\dot{a}_{\perp} >0$ this can then be seen to imply that $\dot{E}_{+}/E_{+} <0$, and so the magnitude of $E_{+}$ is only ever decreasing. As $E_{+}$ is the only independent part of the Riemann curvature that is non-zero, this means that the spacetime curvature must remain finite for all time along every point along every edge. The points where $a_{||}=0$ do not therefore correspond to curvature singularities, but only to conjugate points in the flow of $u^{\mu}$. These constitute caustics in the congruence of time-like curves we are considering, and are somewhat similar to the shell-crossings that can occur in the Lema\^{i}tre-Tolman-Bondi solutions (see e.g. \cite{Bolejko}).

\begin{figure}[t!]
\begin{centering}
\includegraphics[width=6in]{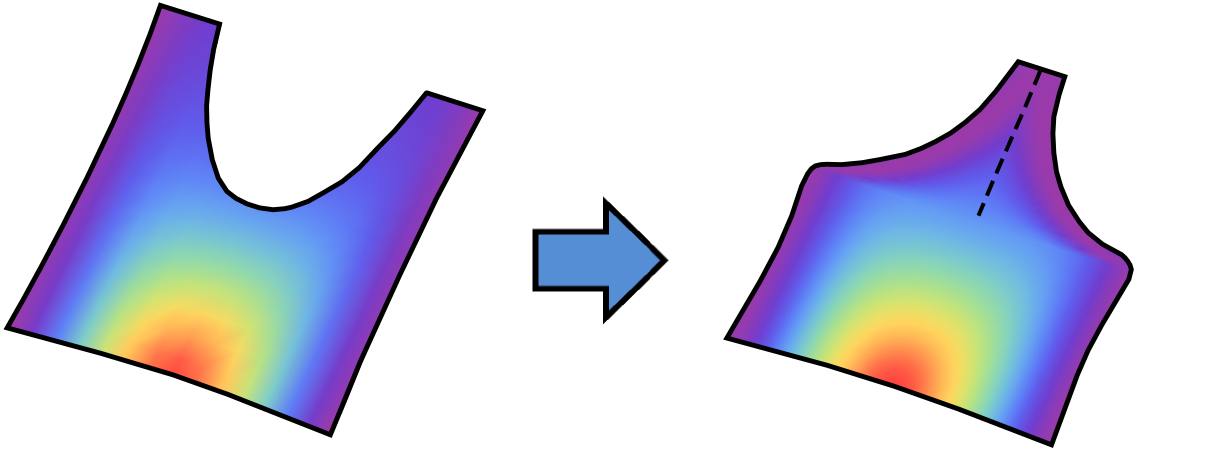}
\par\end{centering}
\caption{A pictorial representation of the identification of the edges of the excluded region (the region above the curved line in the image on left) to form a continuous evolution of the cell edge (image on the right).}
\centering{}\label{cuttingfig}
\end{figure}

It can be seen from Figs. \ref{heat5fig}-\ref{heat600fig} that the form of the profile of $E_{+}$ as a function of $\chi$ maintains its shape as the evolution proceeds, and that it is only the magnitude of $E_{+}$ that decreases as $a_{\perp}$ increases. This behaviour is expected from the evolution equations (\ref{H112})-(\ref{Ep2}), as each point along every edge can be seen to evolve in a similar way, and (once the constraint equations have been satisfied) in a way that is essentially independent of any other {point. The excluded regions in the 5, 8 and 120-cell models} appear to occupy a larger fraction of the edge length than in the {12, 24 and 600-cells}. The reason for this is due to the different functional forms of $(E_{+})_0$ in Figs. \ref{E1fig} and \ref{E2fig}: The former has more points where the magnitude of $E_{+}$ is initially relatively large, while the latter case have more points where the magnitude of $E_{+}$ is initially relatively small. Points where $E_{+}$ is large collapse faster in the direction parallel to the edge, and become excluded at earlier times.

\subsection{Evolution of ${\bf \ell (t)}$}

In performing the integral in Eq. (\ref{edge}) we must integrate along the entire edge of each cell. Before any of the points along the edge reach $a_{||}=0$ this corresponds to a simple integral from $\chi_1$ to $\chi_2$. After some of the points have reached $a_{||}=0$, however, a little more care is required. At the moment when $a_{||}=0$ we have that what were two neighbouring time-like geodesic curves are now no longer space-like separated at all, and can therefore (for the purposes of our integral) be identified. The picture that results is illustrated in Fig. \ref{cuttingfig}. The coordinate separation of the vertices that {demarcate} the ends of the edge is decreased as the excluded region is excised, and the relevant points identified. At this stage the integral in Eq. (\ref{edge}) is then {evaluated over the values} of $\chi$ that have not yet been excluded.

\begin{figure}[t!]
\begin{centering}
\includegraphics[width=6in]{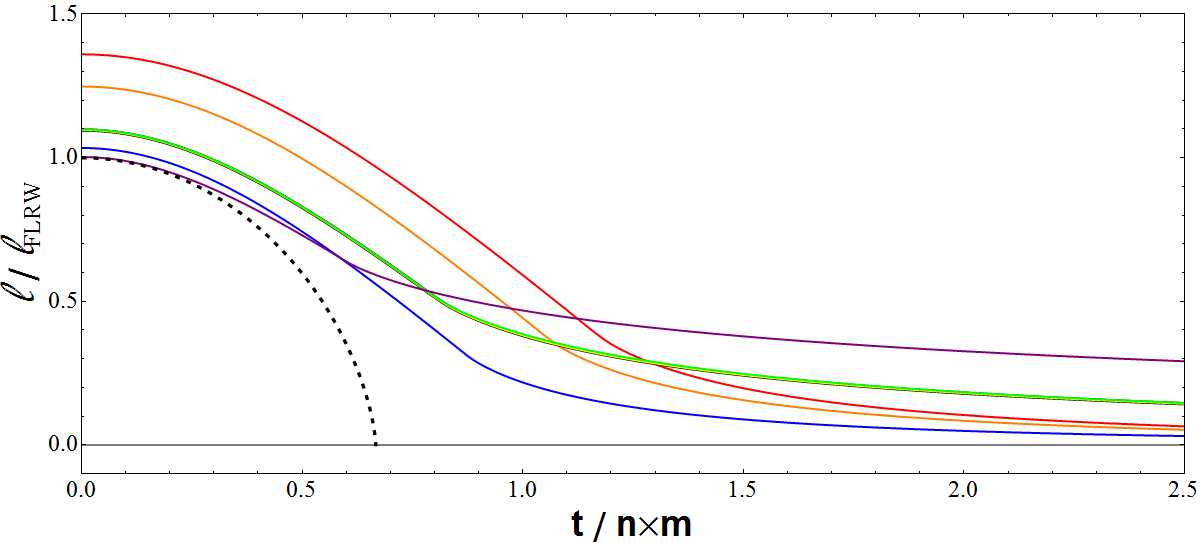}
\par\end{centering}
\caption{The edge length of the six lattices, calculated using Eq. (\ref{ell}), and excluding from the integral points with $a_{||}=0$. The six solid lines denote, from top to bottom at $t=0$, the 5-cell (red), the 8-cell (orange), the 24-cell (green), the 16-cell (yellow), the 120-cell (blue) and the 600-cell (purple). The edge lengths presented here have been normalised by the length they would have in a spatially closed FLRW universe with the same proper mass at $t=0$ (see \cite{lattice1}). Time is presented in units of the total mass in the lattice (that is, the size of the mass in each cell, $m$, multiplied by the number of cells, $n$). The dotted (black) line is a spatially closed FLRW solution with the same total proper mass as each of the lattice models, and is presented for comparison. The yellow and green curves are almost completely indistinguishable by eye in this plot.}
\centering{}\label{edgesfig}
\end{figure}

The results of these considerations for the finite length of an edge in each of our six lattices is displayed graphically in Fig. \ref{edgesfig}. In presenting these data we have chosen to rescale the length of each edge by the length of a similar curve at the maximum of expansion in a closed Robertson-Walker geometry that contains the same total ``proper mass''. A similar curve here is chosen as a curve that subtends the same angle on a hypersurface of constant time, and the proper mass is the mass that an observer who is arbitrarily close to one of the point-like objects would infer on the time-symmetric hypersurface (see Ref. \cite{lattice1} for details). Time in this plot is presented in units of the total proper mass in the spacetime. That is, in units of $n \times m$, where $n$ is the number of cells in the lattice, and $m$ is the mass at the centre of each cell. {  These units allow} us to present the evolution of the edge of a cell in each of our lattices in such a way that they can all be compared to a single FLRW curve that contains the same total proper mass (the dotted black line in Fig. \ref{edgesfig}).

Several interesting features are apparent in Fig. \ref{edgesfig}. Firstly, we can recover the results found in Ref. \cite{lattice1} for the difference in scale of these lattice models compared to FLRW solutions. These numbers are presented in Table \ref{scaletable}, below\footnote{This corrects an error in \cite{lattice1} for the scale of the 8-cell.}. It is apparent that as the number of masses is increased the scale of these models at the time-symmetric hypersurface appears to approach the FLRW value at the maximum of expansion. This pattern can be extended to a smooth limit when considering arbitrarily large numbers of masses, as long as they obey a suitable criterion for being evenly distributed \cite{korGR}. It is also apparent that in the vicinity of the maximum of expansion the evolution of these edge lengths looks very similar to that of a spatially closed FLRW model. This agrees with the findings of the numerical study of the evolution of the 8-cell performed in \cite{BandK8cell}.

Away from the maximum of expansion, however, the evolution of the edge lengths is 
{  widely different from the corresponding} spatially closed FLRW model. There are a number of reasons for this behaviour. Firstly, the evolution of the scale factor $a_{||}$ obeys Eq. (\ref{H112}), which is not the same as the acceleration equation from FLRW cosmology (as can be seen from Fig. \ref{a-fig}). As the evolution of the edge progresses, the curves that show the evolution of the edge in Fig. {\ref{edgesfig}} start to flatten in a similar way to the behaviour of $a_{||}$ in Fig. \ref{a-fig}. This behaviour is expected as the edge lengths and $a_{||}$ are related through Eq. (\ref{edge}). 

What is perhaps less expected is that the second derivative of $\ell (t)$ changes sign during the evolution. This behaviour is not possible for the scale factor $a_{||} (t)$, and is due to the points with $a_{||}\leq 0$ being excluded from the integral in Eq. (\ref{edge}). As already discussed above, we have seen that it is not possible for a curvature singularity to develop at any point along the edge of a cell. Instead, what appears to be happening at late times is that the geometry along the edge is becoming more and more dominated by the regions of Minkowski space that exist near the vertices, as all other parts of the edge collapse to $a_{||}=0$. The expected behaviour in this case is that $\ell(t)$ should flatten out, which is exactly what is observed at late times in Fig. \ref{edgesfig}. The edges then collapse to $\ell=0$ in the {  limit as} $t \rightarrow \infty$.

Once again, the different forms of the initial spatial profile of $E_{+}$ have a discernible effect on the behaviour of the {  lattice edges}. The three lattices that have a non-zero first derivative of $E_{+}$ 
{  at the ends of their edges} all collapse to $\ell \sim 0$ relatively quickly. This fits well into our 
{  picture with Minkowski-like regions} in the vicinity of the vertices dominating the dynamics at late times. In the {case of these} three lattices the extent of the spacetime in the vicinity of the vertices that is close to the geometry of Minkowski space is smaller than in the other three lattices, and therefore the collapse occurs more quickly. For the remaining three lattice, that have a vanishing first derivative of $E_{+}$ at the ends of the cell edges, the value of $\ell (t)$ can be seen to coast at a larger value for a longer time. 

\begin{table}[t!]
\begin{center}
\begin{tabular}{|c|c|}
\hline
\; {\bf Lattice} \; & \; \bf{ $\left. \frac{\ell}{ \ell_{{\rm FLRW}}} \right|_{t=0}$} \; \\
\hline
$5-$cell & $1.360$ \\
$8-$cell & $1.248$ \\
$16-$cell & $1.097$ \\
$24-$cell & $1.099$ \\
$120-$cell & $1.034$ \\
$600-$cell & $1.002$ \\
\hline
\end{tabular}
\end{center}
\caption{{\protect{\textit{The fractional difference in scale between an edge length in each of the lattices, and in a FLRW solution with the same total proper mass, at $t=0$.}}}}
\label{scaletable}
\end{table}

Interestingly, the value of $\ell/\ell_{\rm FLRW}$ at which the coasting occurs does not decrease as the number of cells increases. In fact, the curves corresponding to the 16 and 24-cell lattices are almost indistinguishable (as was already known to be the case for their value at $t=0$ \cite{lattice1}). The curve corresponding to the 600-cell in Fig. \ref{edgesfig} can be seen to collapse the most slowly of all the lattices. This can be understood from Fig. \ref{heat600fig}, as it can be seen that in the 600-cell the region in which $E_{+} \sim 0$ extends much further away from the ends of the edge than is the case in any of the other lattices. The Minkowski-like regions near the vertices are therefore even more dominant in this lattice. Although we have few data points, this behaviour displays what appears to be an interesting lack of convergence with FLRW behaviour as the number of cells in the lattice is increased, {at least in presence of the symmetries considered here}.

\subsection{Hubble Rates and Deceleration Parameters}

\begin{figure*}[t!]
\begin{centering}
\includegraphics[width=6.25in]{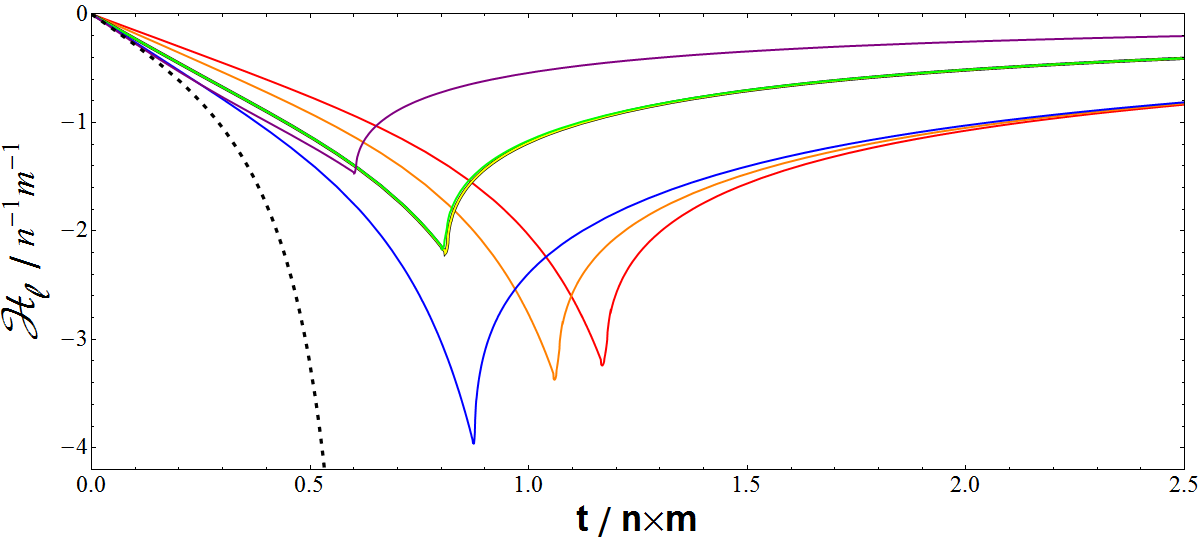}
\par\end{centering}
\caption{The Hubble expansion rate of the edges of the cells, as defined in Eq. (\ref{Heq}). From highest to lowest at early times, the six solid lines correspond to the 5-cell (red), the 8-cell (orange), the 16 and 24-cells (yellow and green), and the 120 and 600-cells (blue and purple). The dotted line is the curve associated with a spatially closed FLRW solution with the same total proper mass. Units of time are again taken to be the total proper mass in the spacetime.}
\centering{}\label{Hubblefig}
\end{figure*}

\begin{figure*}[htb]
\begin{centering}
\includegraphics[width=6.1in]{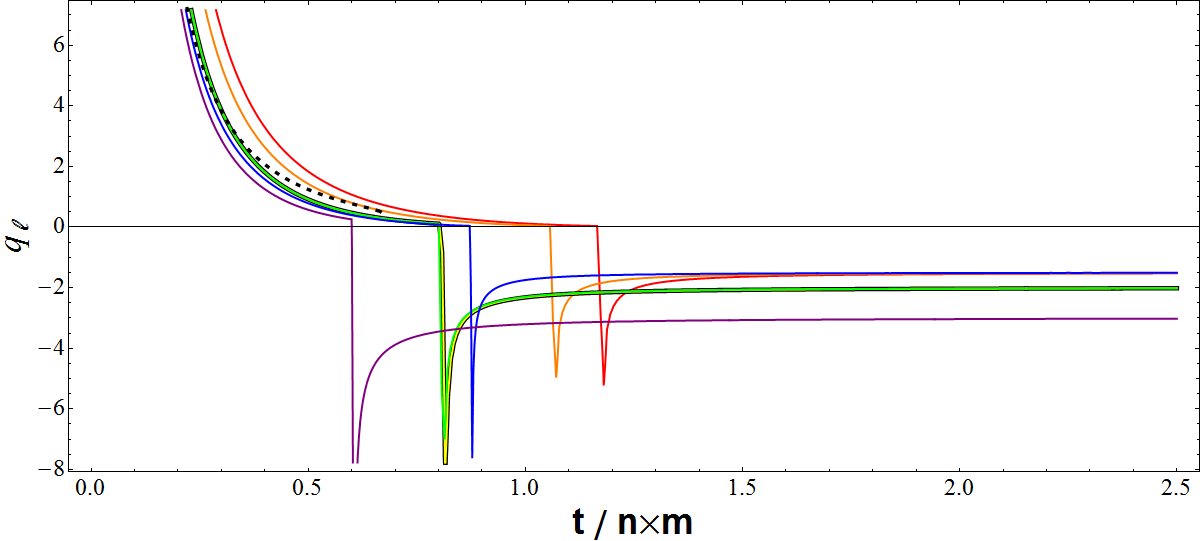}
\par\end{centering}
\caption{The deceleration parameter associated with the expansion rate of the edges of the cells, as defined in Eq. (\ref{qeq}). From highest to lowest at early times, the six solid lines correspond to the 5-cell (red), the 8-cell (orange), the 16 and 24-cells (yellow and green), the 120-cell (blue), and the 600-cell (purple). The dotted line is the curve associated with a spatially closed FLRW solution with the same total proper mass, and time is displayed in units of total proper mass.}
\centering{}\label{qfig}
\end{figure*}

The curves displayed in Fig. \ref{edgesfig} can be used to calculate the effective Hubble rates and deceleration parameters associated with the cell edges in our lattice models. These quantities are defined, respectively, as
\be
\label{Heq}
\mathcal{H}_{\ell} \equiv \frac{\dot{\ell}}{\ell},
\ee
and
\be
\label{qeq}
q_{\ell} \equiv - \frac{\ddot{\ell} \ell}{\dot{\ell}^2}.
\ee
These definitions are made in analogy with the usual quantities that are considered in FLRW cosmology, and the results of calculating them in our lattice models is displayed in Figs. \ref{Hubblefig} and \ref{qfig}. Also displayed in these figures, for reference, is the Hubble rate and deceleration of the spatially closed FLRW solution plotted in Fig. \ref{edgesfig}.

A number of interesting types of behaviour can be observed in Figs. \ref{Hubblefig} and \ref{qfig}. Firstly, it can be seen that there is a kink in the Hubble rate displayed in Fig. \ref{Hubblefig}. In each case this kink occurs at the moment when $a_{\perp}$ first reaches zero at the middle of the edge. The behaviour of the evolution then changes, as might be expected. The effect of this kink on the deceleration parameter is readily apparent in Fig. \ref{qfig} as a sudden change in $q_{\ell}$. Before this point the value of $q_{\ell}$ appears to behave in a qualitatively similar fashion to the FLRW curve. After the kink the behaviour of $q_{\ell}$ is very different, becoming negative, and eventually settling down to a very nearly constant value.

The negative values of $q_{\ell}$ in Fig. \ref{qfig} correspond to acceleration of the length of the edge, a phenomenon that usually requires exotic matter content in FLRW cosmology, and that is posited to have occurred in both the early and late Universe. Here the acceleration does not require any exotic matter, however, and is a consequence of the vacuum dynamics of the solutions to Einstein's equations only. The three lattices with non-contiguous edges all approach the same value of $q_{\ell} \simeq -3/2$ at late times. This corresponds to near power-law evolution, with $\ell \propto t^{-2}$. {In contrast, the 16 and 24-cell models} both approach a value of $q_{\ell} \simeq -2$ at late-times, while the {600-cell model} approaches $q_{\ell} \simeq -3$. This corresponds to power-law evolution of the form $\ell \propto t^{-1}$ and $\ell \propto t^{-\frac{1}{2}}$, respectively. In a spatially flat FLRW solution such behaviour would be taken to correspond to a ``phantom fluid'', with $p < -\rho$.

In Fig. \ref{Hubblefig} it can be seen that at early times, near the time-symmetric hypersurface, there appears to be some convergence towards the FLRW Hubble rate as the number of cells in the lattice is increased. This convergence is short lived, however, as even the {120 and 600-cell models} start to diverge from the FLRW behaviour at relatively early times. The {  locations of the kinks do} not vary in an easily predictable way as the number of cells is increased, either in terms of the time at which they occur, or at their value of $\mathcal{H}_{\ell}$. The three lattices with non-contiguous edges do, however, appear to follow a simple pattern, occurring at earlier times and lower $\mathcal{H}_{\ell}$ as the number of masses is increased.

Finally, there seems to be little or no noticeable convergence to FLRW behaviour in the values of $q_{\ell}$ for our six lattices. That is, although the shape of each curve in Fig. \ref{qfig} appears to initially be similar to the FLRW curve, the precise positions of these curves does not seem to approach the value of the FLRW curve in any obvious way as the number of cells is increased. Similarly, the behaviour after the occurrence of the {kinks does} not seem to follow a simple pattern as the number of masses is increased.

\section{Evolution of the Distance Between Horizons}

\subsection{Position of the Horizons}
\label{sec:horizons}

Methods for locating the positions of black hole horizons are discussed in some detail in Chapter 7 of \cite{baum}. The positions of the horizons will be estimated here by looking for the initial position of the compact orientable 2-surfaces that are Marginally Outer Trapped Surfaces (MOTS), and then following the evolution of these surfaces at the points where they intersect with our LRS curves. These surfaces initially have future directed outward null normals, $k^{\mu}$, with vanishing expansion, $k^{\mu}_{\phantom{\mu} ; \mu}=0$, and hence are non-expanding horizons \cite{Ash}. If we align $e_1^{\phantom{1} \mu}$ with the space-like direction normal to the horizon, then the vanishing expansion gives
\be
e_{0 \mu ; \nu} g^{\mu \nu} + e_{1 \mu ; \nu} g^{\mu \nu}=0.
\ee
The first term in this equation is the trace of the extrinsic curvature of the initial hypersurface, which vanishes due to time symmetry. The second term is the expansion of ${e}_1^{\phantom{1} \mu}$ (since $\dot{u}^{\mu}=0$), and so we have that the MOTS are minimal surfaces in the time-symmetric hypersurface \cite{Gibbons}.

The initial positions of these horizons were estimated in \cite{lattice1}, by rotating the lattice until one of the masses appears at $\chi=0$, and then looking for the sphere of constant $\chi$ that is centred around this point and has the minimal surface area. Here, however, we need to be more accurate, so we proceed by recognising that the MOTS are totally geodesic. That this is true can be seen from Raychaudhuri's equation applied to $k^{\mu}$, which gives
\be
\label{nullRay}
\mathcal{L}_k \tilde{\Theta} = -\frac{1}{2} \tilde{\Theta}^2 - \tilde{\sigma}^{\mu \nu}\tilde{\sigma}_{\mu \nu} + \tilde{\omega}^{\mu \nu} \tilde{\omega}_{\mu \nu} - R_{\mu \nu} k^{\mu} k^{\nu},
\ee
where $\tilde{\Theta}$, $\tilde{\sigma}_{\mu \nu}$ and $\tilde{\omega}_{\mu \nu}$ denote the expansion, shear and vorticity of the null geodesics to which $k^{\mu}$ is tangent. We have $\tilde{\omega}_{\mu \nu}=0$ as these null geodesics are surface forming, and $R_{\mu \nu}=0$ as our spacetime is vacuum. The conditions that the generators of the horizon have vanishing expansion and are initially null then gives $\tilde{\Theta}=0$ and $\mathcal{L}_k \tilde{\Theta}\vert_{t=0} = 0$. From Eq. (\ref{nullRay}) we then have that $\tilde{\sigma}^{\mu \nu}\tilde{\sigma}_{\mu \nu}\vert_{t=0} =0$, which implies that $\tilde{\sigma}_{\mu \nu}\vert_{t=0}=0$. As the shear and expansion of $u^{\mu}$ are initially zero, i.e. $\sigma_{\mu \nu}\vert_{t=0}=0=\Theta \vert_{t=0}$, this is sufficient to prove the MOTS on the time-symmetric hypersurface are extrinsically flat, and hence are totally geodesic with indeterminate lines of curvature \cite{Eis}.

It is known that any $n$-dimensional space that admits an ($n$-$1$)-dimensional surface with indeterminate lines of curvature, and that has constant mean curvature, has a Ricci principal direction that coincides with the normal to that surface, at all points on that surface \cite{Eis}. From Eq. (\ref{Eab}) it can then be seen that the space-like normal to the intersection of the MOTS with the initial hypersurface, $e_1^{\phantom{1} \mu}$, is a principal direction of $E_{\alpha \beta}$. This means that in an orthonormal tetrad that contains $e_1^{\phantom{1} \mu}$ we have $E_{12}=E_{13}=0$ at all points on the horizon. It can then be seen from Eq. (\ref{con7}) that we have
\be
 {\bf e}_{1} (E^{11}) = \frac{3}{2} \theta_1 E^{11} +n_{23} \left( E^3_{\phantom{3}3}-  E^2_{\phantom{2}2} \right),
\ee
where $\theta_1=e_{1 \mu ; \nu} h^{\mu \nu}=2 a_1$ is the expansion of ${e}_1^{\phantom{1} \mu}$ in the space-like direction along which it points. 

If we now recall that on MOTS the tangent vector along an outgoing null geodesic satisfies $k^{\mu}_{\phantom{\mu} ; \mu} =0$, and that $u^{\mu}_{\phantom{\mu} ; \mu}=0$ on our intial hypersurface, then the location of the horizon at LRS points can be seen to be given by
\be
\label{initialhorizon}
 {\bf e}_{1} (E^{11}) =0.
\ee
That is, the initial position of the horizons are at local extrema of $E_{11}$ along any outgoing LRS null curve.

To find the position of the horizon along our LRS curves at later times we can again make use of Eq. (\ref{nullRay}). If $k^{\mu}$ in this equation are the set of outward directed null curves that pass through the MOTS on the time-symmetric initial hypersurface, then at $t=0$ we have
\be
\tilde{\Theta} \vert_{t=0} =0.
\ee
Now consider these null geodesics to lie within a neighbourhood of a set of space-like LRS curves. We then have $\tilde{\sigma}_{\mu \nu} = 0 =\tilde{\omega}_{\mu \nu}$, and Eq. (\ref{nullRay}) becomes
\be
\label{nullRay2}
\mathcal{L}_k \tilde{\Theta} + \frac{1}{2} \tilde{\Theta}^2 =0.
\ee
This equation is to be solved with the initial condition $\tilde{\Theta}=0$, which admits as its only solution $\tilde{\Theta}=0$ along the entire flow of $k^{\mu}$. This means that every point along an outgoing null geodesic that passes through the horizon at the initial time, and is directed along a LRS curve, has the property $\tilde{\Theta}=0$, and is therefore part of a marginally trapped surface. 

These results allow us to locate the positions of the MOTS along any LRS curve throughout the entire evolution of the spacetime by simply propagating out a null geodesic from the initial point specified by Eq. (\ref{initialhorizon}). That is, these curves are generators of the horizon.

\subsection{Curves Through Cell Faces}
\label{sec:faces}

\begin{figure}[t!]
\begin{centering}
\includegraphics[width=6in]{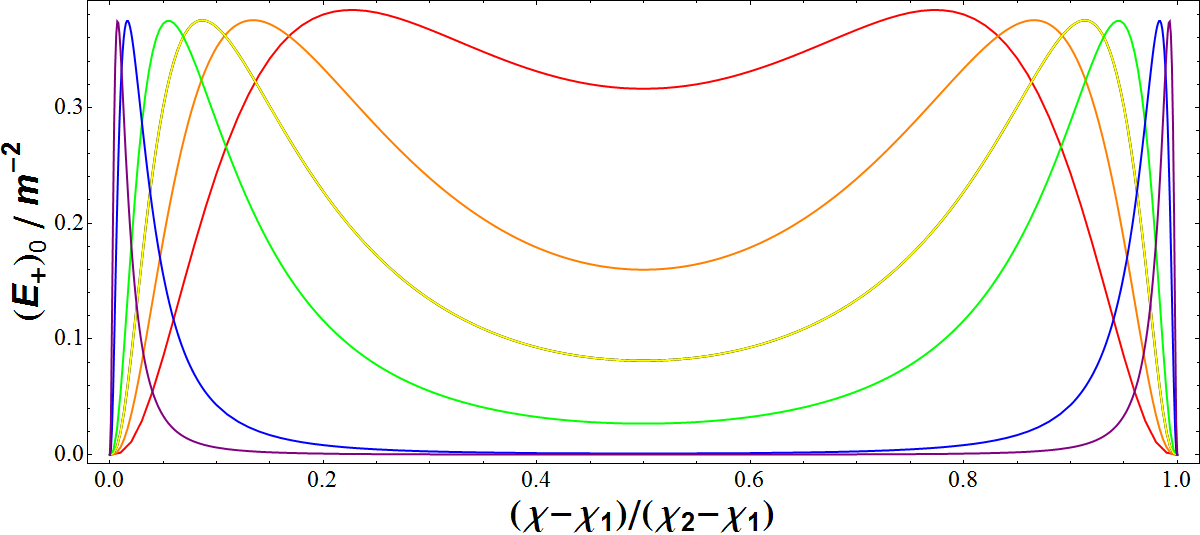}
\par\end{centering}
\caption{The value of $(E_{+})_0$ along a curve that connects cell centres and passes through the centre of a face, in units of $m^{-2}$. At the mid-point the upper curve (red) corresponds to the 5-cell, and then in descending order are the 8-cell (orange), the 16-cell (yellow), the 24-cell (green), the 120-cell (blue), and the 600-cell (purple).}
\centering{}\label{E1horizonsfig}
\end{figure}

As in the discussion of the evolution of cell edges presented in Sec. \ref{sec:edges}, let us first consider the evolution of $E_{+}$, which acts as a source in Eqs. (\ref{H112}) and (\ref{Hp2}). The initial profile of $E_{+}$ along the LRS curves that connect horizons, and that pass through the centres of cell faces, are shown in Fig. \ref{E1horizonsfig}. The coordinate position $\chi_1$ and $\chi_2$ correspond to the locations of the centres of two neighbouring cells, when the lattice has been rotated so that the curve that connects two cell centres through a cell face has $\theta=$constant and $\phi=$constant.

\begin{figure}
\centering
  \subfloat[$E_{+}$ between horizons in the 5-cell.]{\label{heat5fig2}
    \includegraphics[width=4in]{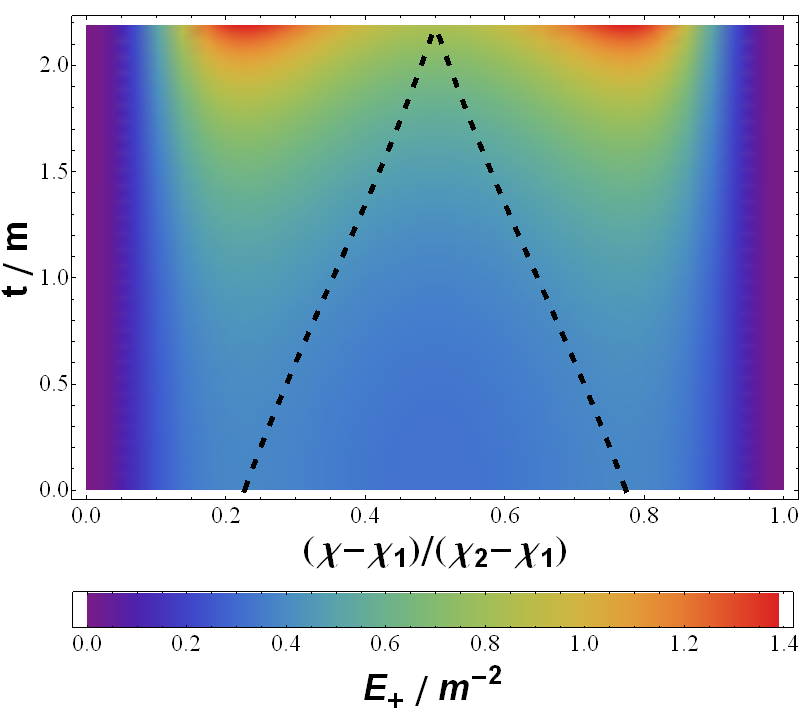}}\qquad\qquad
  \subfloat[$E_{+}$ between horizons in the 8-cell.]{\label{heat8fig2}
    \includegraphics[width=4in]{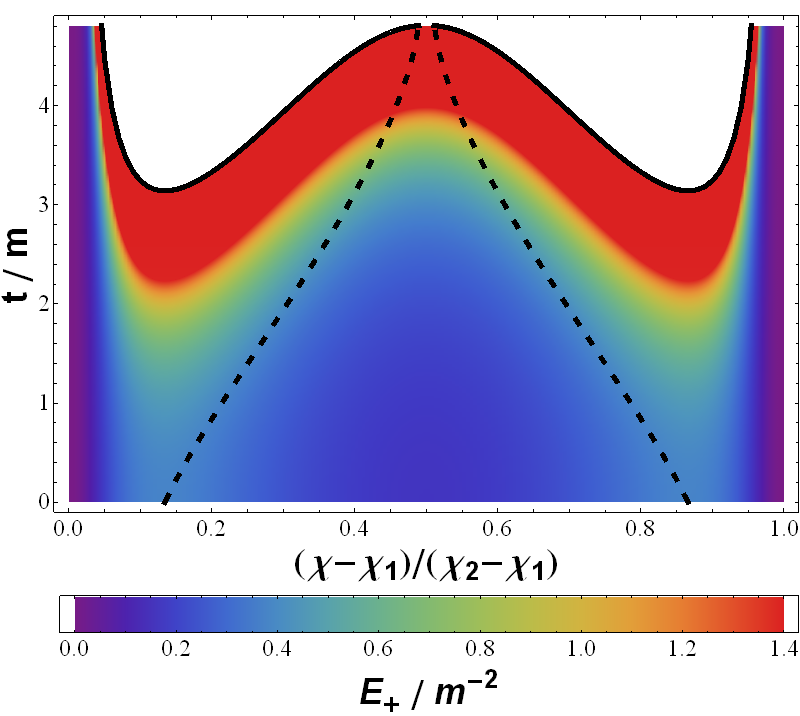}}
\caption{Spacetime diagrams of $E_{+}$ along a curve that connects horizons, and that passes through the centre of a cell face. Excluded regions, above the black line, have passed the point at which a curvature singularity has formed. The dotted lines show the positions of the horizons. Red areas denote regions at or above the upper limit on the accompanying scale.}
\end{figure}

\begin{figure}
\centering
  \subfloat[$E_{+}$ between horizons in the 16-cell.]{\label{heat16fig2}
    \includegraphics[width=4in]{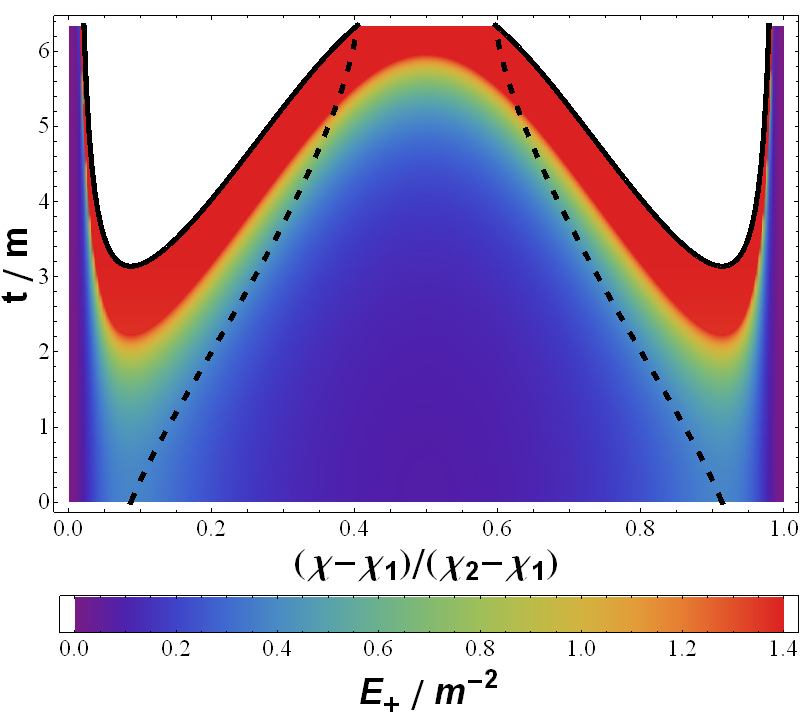}}\qquad\qquad
  \subfloat[$E_{+}$ between horizons in the 24-cell.]{\label{heat24fig2}
  \includegraphics[width=4in]{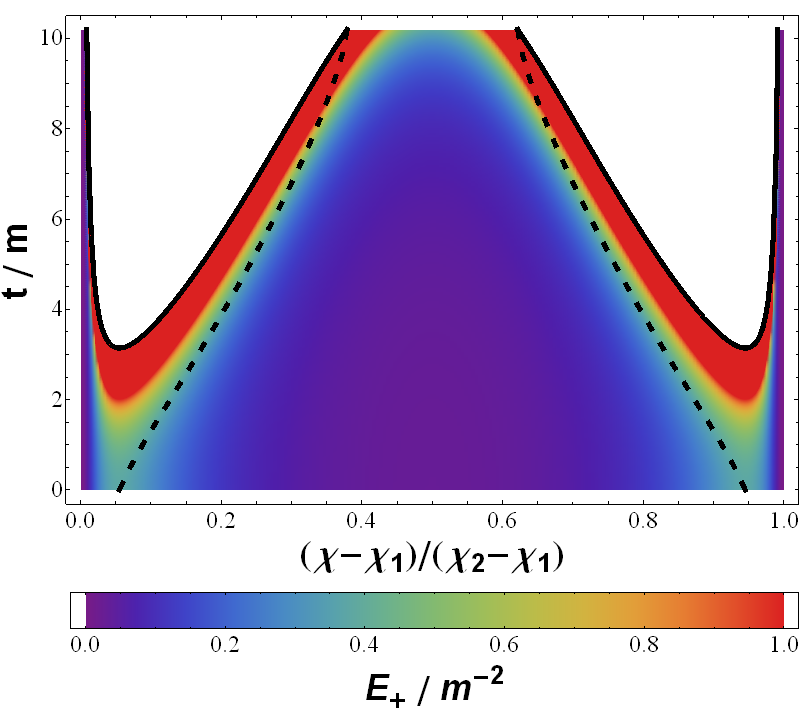}}
\caption{Spacetime diagrams of $E_{+}$ along a curve that connects horizons, and that passes through the centre of a cell face. Excluded regions, above the black line, have passed the point at which a curvature singularity has formed. The dotted lines show the positions of the horizons. Red areas denote regions at or above the upper limit on the accompanying scale.}
\end{figure}

\begin{figure}
\centering
  \subfloat[$E_{+}$ between horizons in the 120-cell.]{\label{heat120fig2}
    \includegraphics[width=4in]{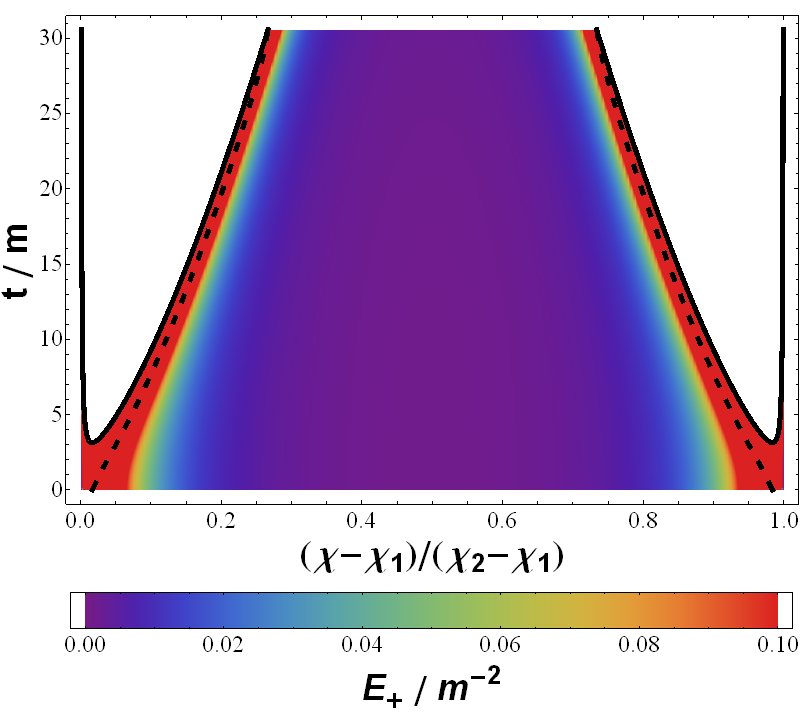}}\qquad\qquad
  \subfloat[$E_{+}$ between horizons in the 600-cell.]{\label{heat600fig2}
  \includegraphics[width=4in]{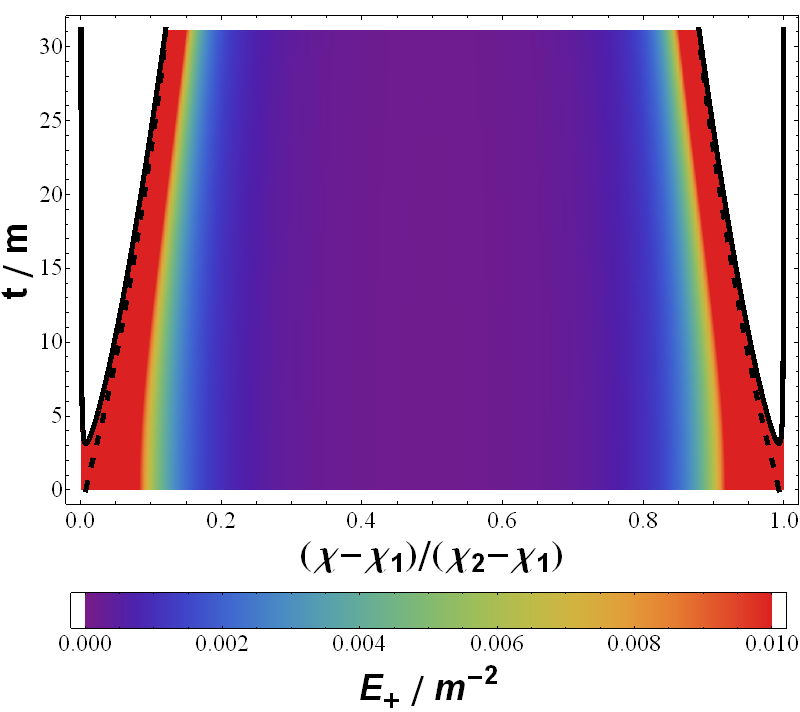}}
  \caption{Spacetime diagrams of $E_{+}$ along a curve that connects horizons, and that passes through the centre of a cell face. Excluded regions, above the black line, have passed the point at which a curvature singularity has formed. The dotted lines show the positions of the horizons. Red areas denote regions at or above the upper limit on the accompanying scale.}
\end{figure}

\begin{figure}[t!]
\begin{centering}
\includegraphics[width=4.6in]{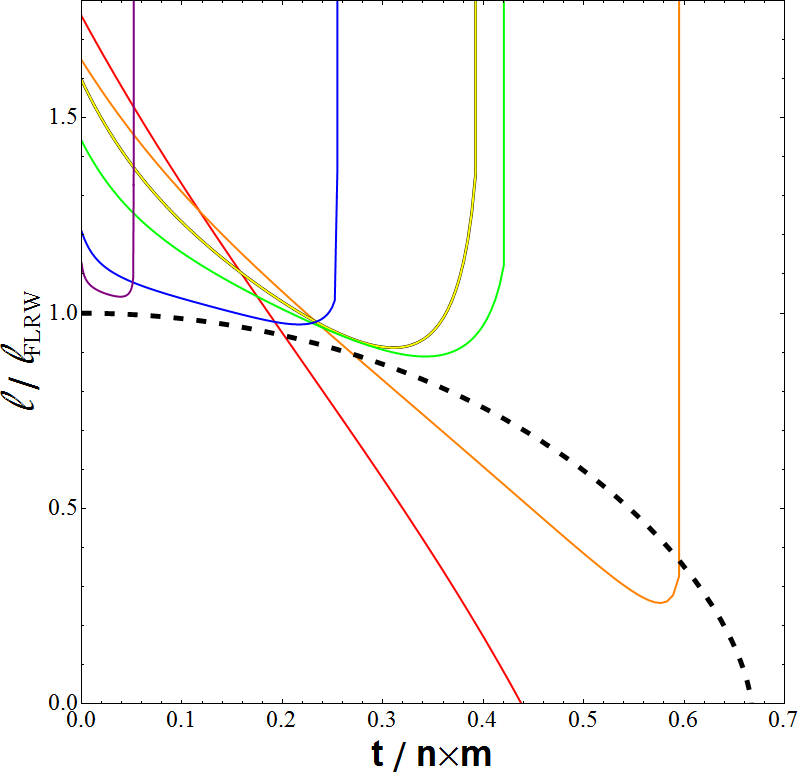}
\par\end{centering}
\caption{The distance between horizons in the six lattices, calculated along a curve that passes through a cell face using Eq. (\ref{ell}). The six solid lines denote, from top to bottom at $t=0$, the 5-cell (red), the 8-cell (orange), the 16-cell (yellow), the 24-cell (green), the 120-cell (blue), and the 600-cell (purple). The distances presented here have been normalised by the length of a curve at $t=0$ in a spatially closed FLRW universe with the same proper mass that initially subtends the same angle on the hypersphere. As in Fig. \ref{edgesfig}, time is presented here in units of the total mass in the lattice. The dotted (black) line is a spatially closed FLRW solution, for comparison.}
\centering{}\label{horizonsfig}
\end{figure}

Unlike in Figs. \ref{E1fig} and \ref{E2fig}, the form of each of these curves, for each of our six lattices, look similar. All curves {dip towards} $E_{+}=0$ at the cell centres, and each has a local minimum at the location of the cell face, and a local maximum somewhere between the cell face and the cell centre. From the discussion in Sec. \ref{sec:horizons} it is apparent that the initial position of the horizon around each mass is located at the maxima of the curves displayed in Fig. \ref{E1horizonsfig}. The particular numerical values of $\chi$ at the location of the horizons, determined in this way, is in good agreement with the estimates made in \cite{lattice1}.

The evolution of $E_{+}$ along the $u^{\mu}$ that are geodesic and initially normal to our time-symmetric initial hypersurface are shown in Figs. \ref{heat5fig2}-\ref{heat600fig2}. Superimposed on each of these plots is the position of the horizons, as determined using the method outlined in Sec. \ref{sec:horizons}. In each case we choose to stop the evolution when the horizons of neighbouring cells merge, or when the spacetime becomes singular at the location of the horizon. Such singularities will eventually occur at all points along the {curves we are} considering, except at the cell centres. This can be seen from Eqs. (\ref{Hp2}) and (\ref{Ep2}), as for positive $E_{+}$ it is the case that $a_{\perp}$ will always collapse to zero in finite time, at which point $E_{+}$ diverges. The cell centres are exempt from this behaviour as {at these locations we} have $E_{+}=0$ for all time, and they are hence locally Riemann flat.

The evolution of $E_{+}$ in the 5-cell, displayed in Fig. \ref{heat5fig2}, can be seen to be an anomalous case as it is the only lattice in which the horizons merge. In all other lattices the spacetime in the vicinity of the horizon becomes singular before this can occur. We note that this does not preclude the possibility of the formation of a second horizon that encompasses both masses as the evolution proceeds, only that the two inner horizons never seem to merge. The 5-cell is also anomalous as it is the only case in which no point in the spacetime becomes singular before we stop the evolution (although it can be seen that $E_{+}$ is starting to become large in some places).

\begin{table}[t!]
\begin{center}
\begin{tabular}{|c|c|c|}
\hline
\; {\bf Lattice} \; & \; \bf{ $\left.  \frac{\ell}{ \ell_{{\rm FLRW}}} \right|_{t=0}$} \; & \;  $\frac{t_{\rm divergence}}{n \times m}$    \; \\
\hline
$5-$cell & $1.761$ &  $-$\\
$8-$cell & $1.648$ & $0.601$\\
$16-$cell & $1.597$ & $0.396$\\
$24-$cell & $1.441$ & $0.425$\\
$120-$cell & $1.210$ & $0.255$\\
$600-$cell & $1.128$ & $0.052$\\
\hline
\end{tabular}
\end{center}
\caption{{\protect{\textit{The fractional difference in scale between the horizon separation distance in each of the lattices and in a FLRW solution with the same total proper mass at $t=0$, and the time at which the horizon separation distance diverges.}}}}
\label{scaletable2}
\end{table}

The corresponding evolution of the proper distance between horizons is calculated using Eq. (\ref{edge}), and is displayed graphically in Fig. \ref{horizonsfig}. The length of the curve that connects the horizons has, in this case, been scaled by the length of a curve in a spatially closed FLRW solution with the same total proper mass that initially subtends the same angle on the hypersphere (we note that because of the motion of the horizons, in terms of the spatial coordinate $\chi$, the angle subtended by this curve in the lattice models does not stay constant, as it does in FLRW). The units of time in this figure have again been chosen to be the total proper mass of the lattice, in order that all lattices can be compared to a single FLRW solution.

It can be seen that the curve {  corresponding to the} 5-cell in Fig. \ref{horizonsfig} does not diverge at finite time. This is due to the horizons merging in this case. Conversely, the divergence of the horizon distance in the {8, 16, 24, 120 and the 600-cell models} is a consequence of the geometry becoming singular in the vicinity of the horizons. When {this occurs,} $a_{\perp} \rightarrow 0$ and $a_{||} \rightarrow \infty$. It is the divergence of the scale factor in the direction along the curve that causes the proper distance of the entire curve to blow up. The precise time at which these divergences occur is displayed in Table \ref{scaletable2}, along with the scale of the curve as a fraction of its {FLRW value} at $t=0$.

{We note that} these results appear to be in agreement with the numerical results found for the 8-cell by Bentivegna and Korzy\'{n}ski in \cite{BandK8cell}. As is the case in their numerical study, we also find that the evolution of the horizon separation distance at $t=0$ has a non-zero first derivative, which we attribute to the cause identified in \cite{BandK8cell}: that there are two marginally trapped surfaces in the full evolution of the spacetime, whose trajectories intersect each other at the moment of time symmetry. Our results extend this conclusion to the other regular lattices displayed in Fig. \ref{horizonsfig}.

As for the edge lengths calculated in Sec. \ref{sec:edges}, it can be seen that FLRW behaviour is again not always approached as the number of masses in the lattice becomes large. This is particularly true of the time at which the horizon separation distance diverges, which decreases (as a fraction of the age of the corresponding FLRW solution) as the number of masses is increased. This suggests that the big crunch is reached by the spacetime in the vicinity of black holes much faster if the matter in the universe is split up into a larger number of particles. Finally, we note that while the behaviour of the horizon separation distance is similar in the cases of the 16 and 24-cell lattices, it is not as close as in the evolution of the edge lengths.

\subsection{Curves Through Cell Corners}

As well as considering the length of curves that connect horizons and that pass through cell centres, as was done in Sec. \ref{sec:faces}, we can also consider the evolution of a curve that passes through a cell centre and that extends through a vertex of the same cell. Such curves can be used to determine the distance between the horizon and  a vertex, and will be referred to as ``diagonals'' in what follows. These curves exhibit local rotational symmetry, in the same way as the curves considered in Sec. \ref{sec:faces}, and so can be described using the same basic equations. Again we rotate so that the curves in which we are interested are at constant $\theta$ and $\phi$. Here we label the cell centre as being at $\chi_1$, and the vertex in which we are interested as being at $\chi_2$.

\begin{figure}[t!]
\begin{centering}
\includegraphics[width=4.1in]{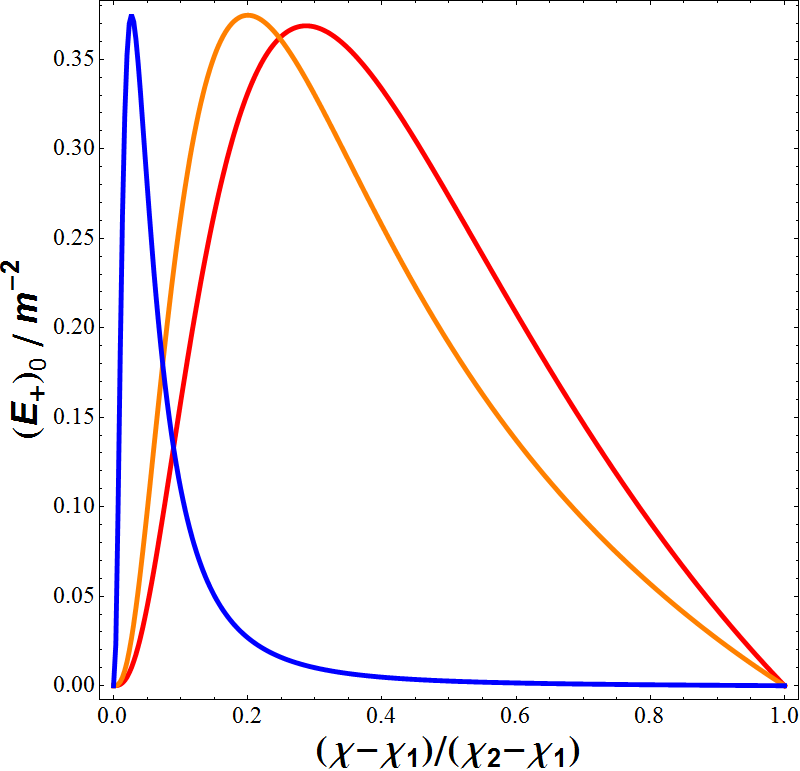}
\par\end{centering}
\caption{The value of $(E_{+})_0$ along a curve that connects a cell centre at $\chi_1$ to a vertex at $\chi_2$. At their peak values, from the left to right, these curves correspond to the 120-cell (blue), the 8-cell (orange), and the 5-cell (red). Each of these curves has non-zero first derivative at $\chi_2$.}
\centering{}\label{E1diagonalfig1}
\end{figure}

\begin{figure}[t!]
\begin{centering}
\includegraphics[width=4.1in]{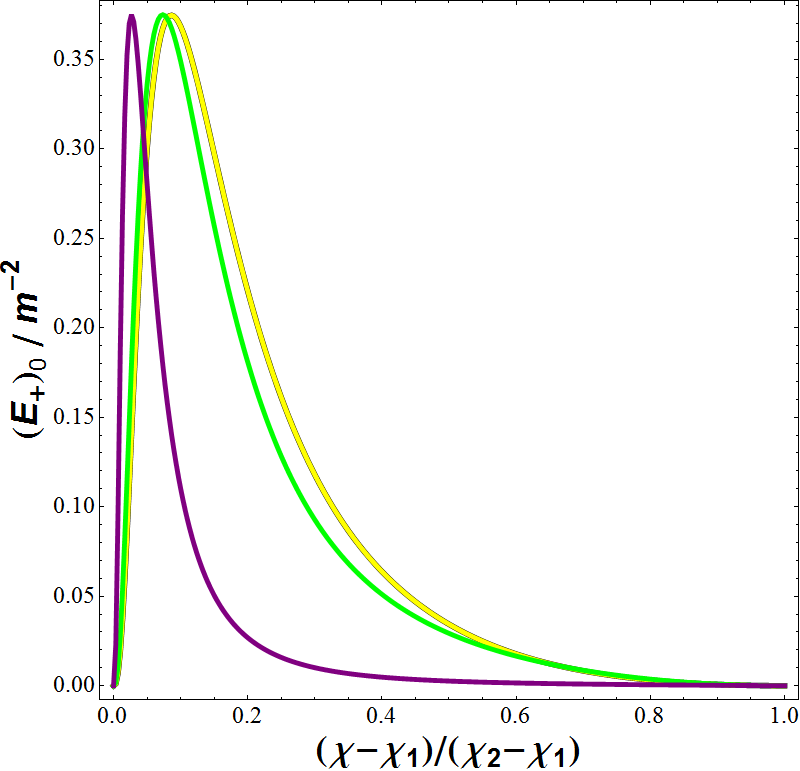}
\par\end{centering}
\caption{The value of $(E_{+})_0$ along a curve that connects a cell centre at $\chi_1$ to a vertex at $\chi_2$. At their peak values, from the left to right, these curves correspond to the 600-cell (purple), the 24-cell (green), and the 16-cell (yellow). Each of these curves has vanishing first derivative at $\chi_2$.}
\centering{}\label{E1diagonalfig2}
\end{figure}

As before, $E_{+}$ acts {as a source} term in Eqs. (\ref{H112}) and (\ref{Hp2}), and is hence of great importance for the evolution of the length of the diagonals. We plot the initial profile of $E_{+}$ along a diagonal in Figs. \ref{E1diagonalfig1} and \ref{E1diagonalfig2}. The former of these shows the curves that have a non-zero gradient at the vertex, while the latter shows those that have a vanishing gradient at the vertex. These two cases correspond to two qualitatively different profiles, as was the case along the edges described in Sec. \ref{sec:edges}. The magnitude of $E_{+}$ at its maximum in Figs. \ref{E1diagonalfig1} and \ref{E1diagonalfig2} can be seen to be similar to the magnitude of the maximum in Fig. \ref{E1horizonsfig}, along the curves that connect the horizons through the centre of a cell face.

The reason for the non-zero {gradients in the cases of the 5, 8 and the 120-cell models} is that the extension of the diagonal out of the cell we are considering corresponds to an edge that is shared by neighbouring cells. As the value of $E_{+}$ is known to be negative along an edge in {  these lattices, and to have non-zero gradients at the ends of the} edge, they must also have non-zero gradients in Fig. \ref{E1diagonalfig1}. For the {16, 24 and the 600-cell models}, however, the extension of the diagonal out of the initial cell corresponds to a diagonal in a neighbouring cell. Because of this we must have symmetry about the vertex, and hence a vanishing first derivative. The consequences of these different profiles for the evolution of the diagonals will become clear in what follow, but we can already note that the area of low curvature around the vertex is large if the derivative of $E_{+}$ is zero at that vertex.

\begin{figure}
\centering
  \subfloat[$E_{+}$ along a diagonal in the 5-cell.]{\label{heat5fig3}
    \includegraphics[width=3.9in]{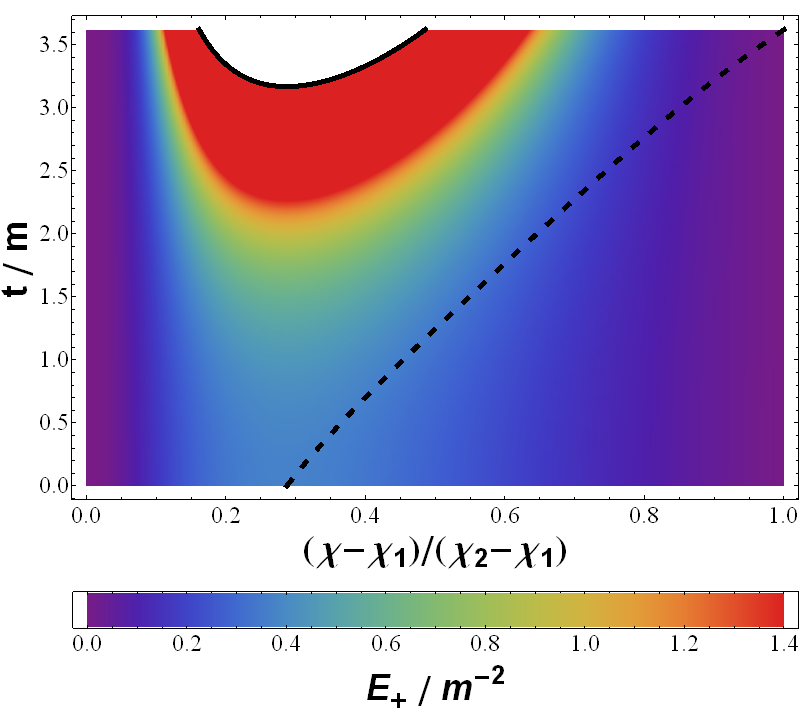}}\qquad\qquad
  \subfloat[$E_{+}$ along a diagonal in the 8-cell.]{\label{heat8fig3}
    \includegraphics[width=3.9in]{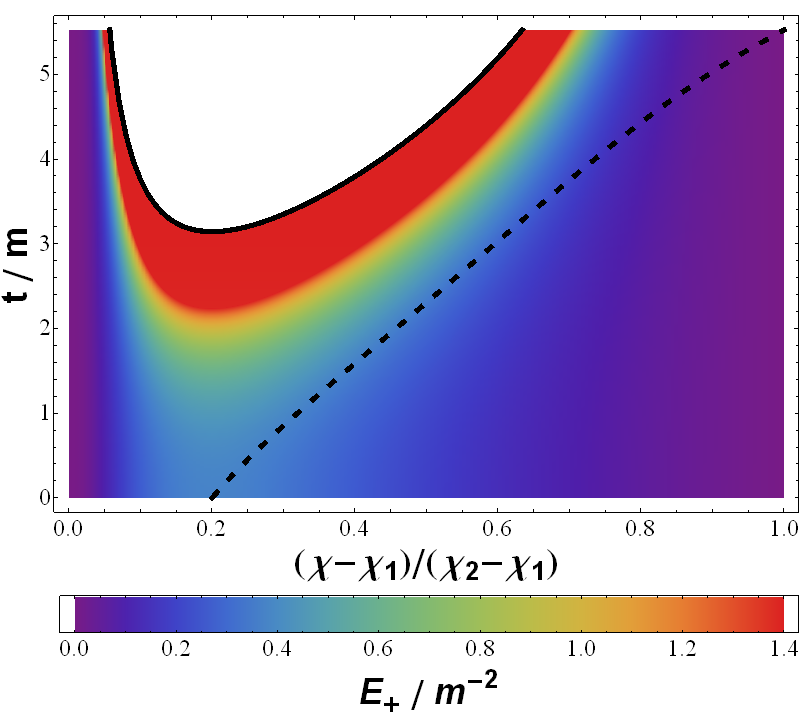}}
 \caption{Spacetime diagrams of the value of $E_{+}$ along a curve that connects the centre of a cell and a vertex, in each of the six lattices. Excluded regions, above the black line, have passed the point at which a curvature singularity has formed. The dotted lines show the position of the horizon. Red areas denote regions at or above the upper limit on the accompanying scale.}
\end{figure}

\begin{figure}
\centering
  \subfloat[$E_{+}$ along a diagonal in the 16-cell.]{\label{heat16fig3}
    \includegraphics[width=3.9in]{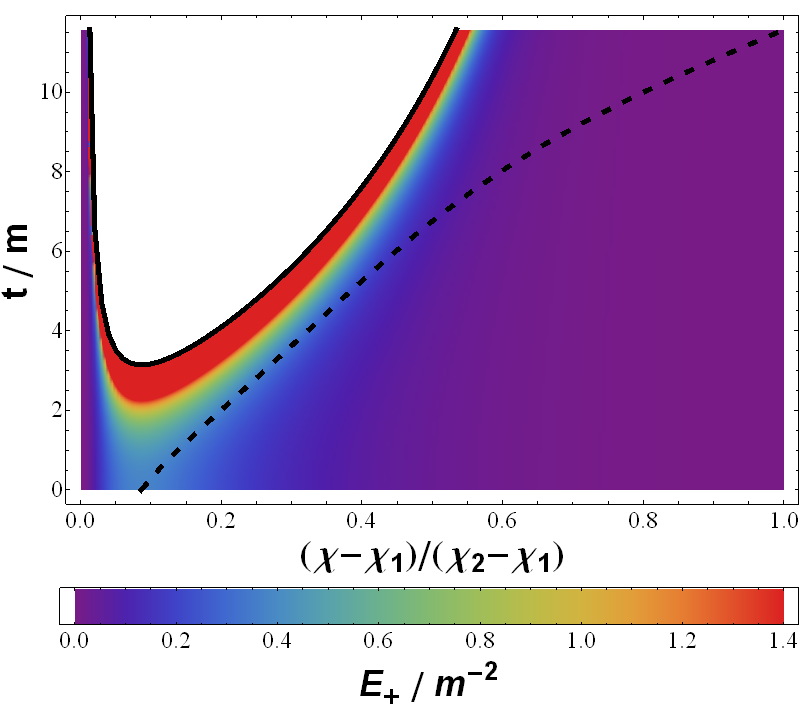}}\qquad\qquad
  \subfloat[$E_{+}$ along a diagonal in the 24-cell.]{\label{heat24fig3}
  \includegraphics[width=3.9in]{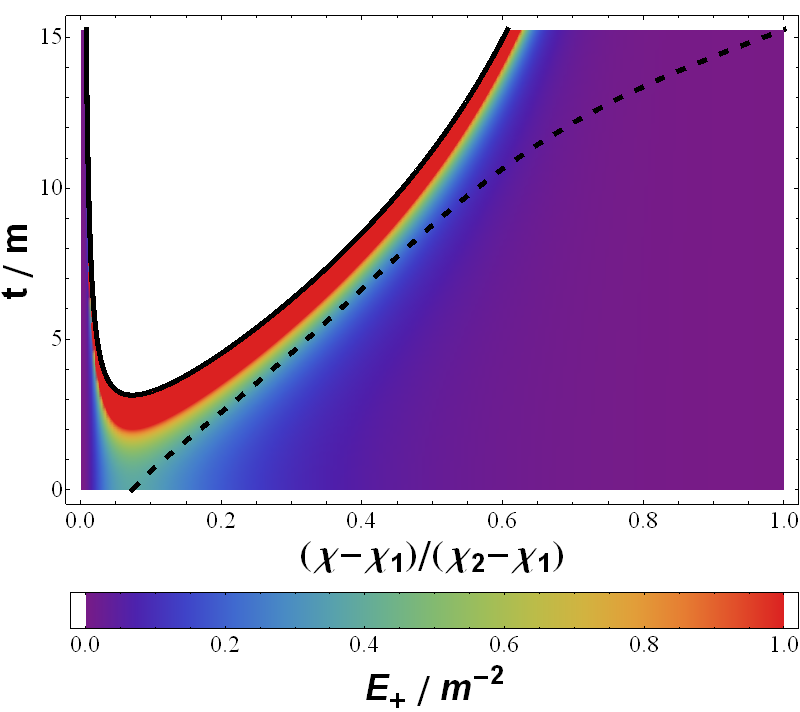}}
 \caption{Spacetime diagrams of the value of $E_{+}$ along a curve that connects the centre of a cell and a vertex, in each of the six lattices. Excluded regions, above the black line, have passed the point at which a curvature singularity has formed. The dotted lines show the position of the horizon. Red areas denote regions at or above the upper limit on the accompanying scale.}
\end{figure}

\begin{figure}
\centering
  \subfloat[$E_{+}$ along a diagonal in the 120-cell.]{\label{heat120fig3}
    \includegraphics[width=3.9in]{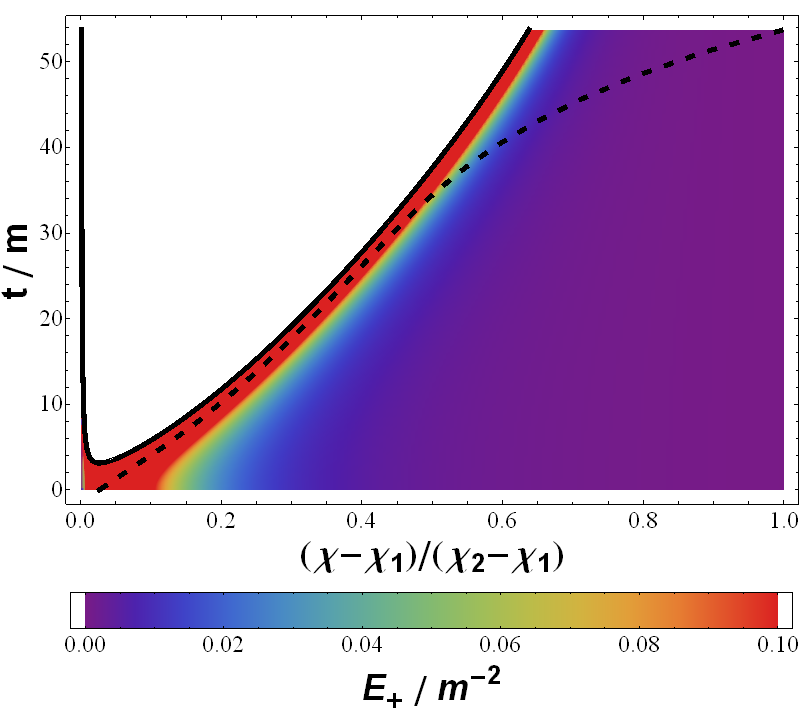}}\qquad\qquad
  \subfloat[$E_{+}$ along a diagonal in the 600-cell.]{\label{heat600fig3}
  \includegraphics[width=3.9in]{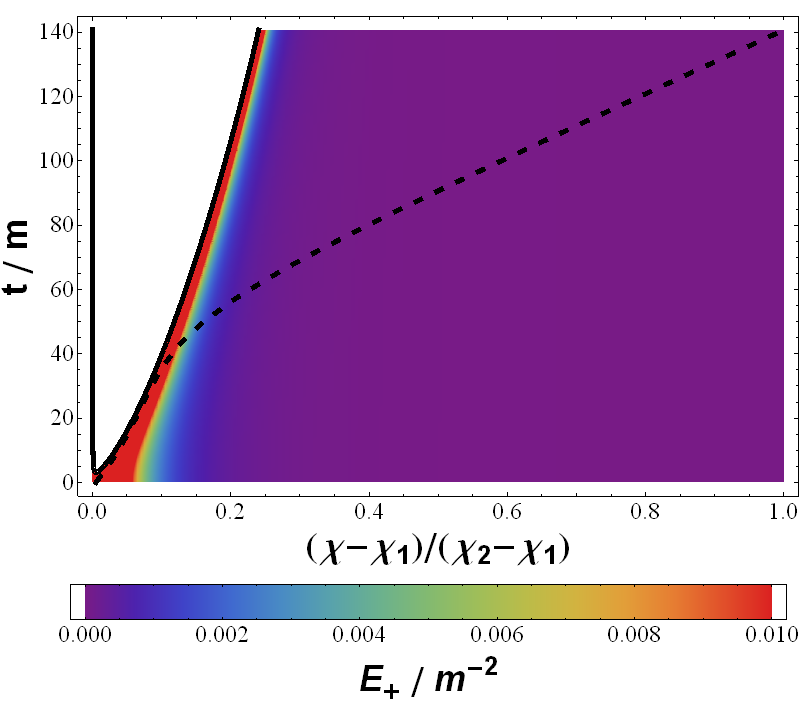}}
  \caption{Spacetime diagrams of the value of $E_{+}$ along a curve that connects the centre of a cell and a vertex, in each of the six lattices. Excluded regions, above the black line, have passed the point at which a curvature singularity has formed. The dotted lines show the position of the horizon. Red areas denote regions at or above the upper limit on the accompanying scale.}
\end{figure}

The evolution of $E_{+}$ along the diagonals is shown in Figs. \ref{heat5fig3}-\ref{heat600fig3}. As with the profiles displayed in Figs. \ref{heat5fig2}-\ref{heat600fig2}, we find here that curvature singularities develop at finite times. Unlike the case of the curves that connect horizons by passing through the centre of a face, however, it can be seen that along a diagonal the horizon will always leave the cell before it encounters a singularity. In the cases of the {16, 24 and 600-cell models} this corresponds to the horizons of neighbouring masses merging before the final singularity. This behaviour can be understood by recognising that the spacetime in the vicinity of a vertex has very low curvature in every lattice. This is not the case for 
{  the regions around the centres of cell faces}. The lower curvature means that the final singularity is arrived at after a much larger {  interval of time}, and so the horizon has more time to make its way to the edge of the cell.

\begin{figure}[t!]
\begin{centering}
\includegraphics[width=4.6in]{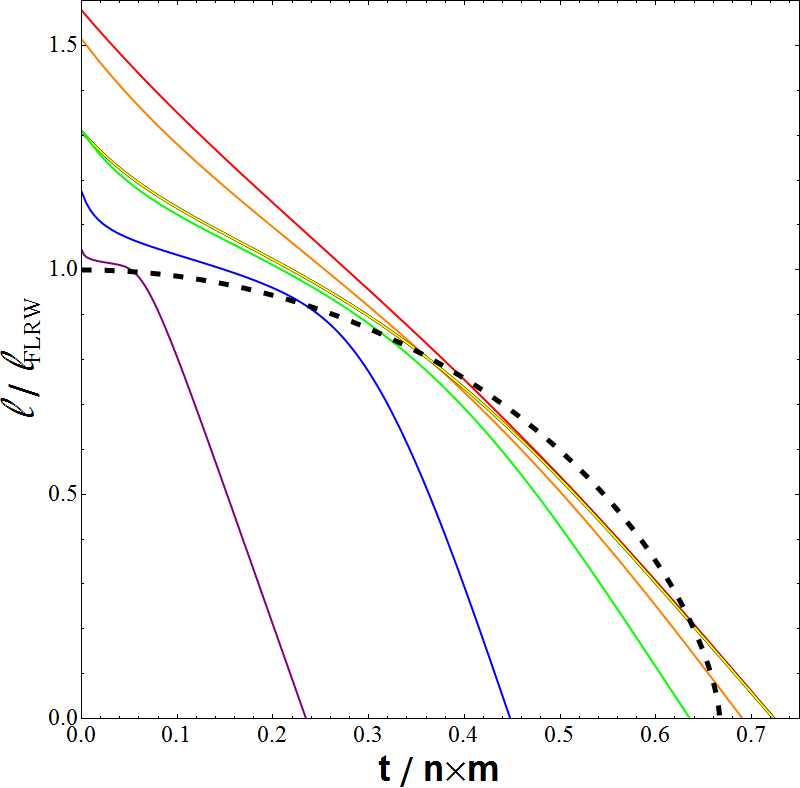}
\par\end{centering}
\caption{The distance between the horizon and a vertex in the six lattices, calculated along a curve that passes through the centre of the cell. The six solid lines denote, from top to bottom at $t=0$, the 5-cell (red), the 8-cell (orange), the 24-cell (green), the 16-cell (yellow), the 120-cell (blue), and the 600-cell (purple). The distances presented here have been normalised by the length of a curve at $t=0$ in a spatially closed FLRW universe with the same total proper mass that initially subtends the same angle on the hypersphere. Time is in units of the total mass in the lattice, and the dotted (black) line is the evolution of the scale factor in a spatially closed FLRW solution with the same total proper mass, for comparison.}
\centering{}\label{diagonalfig}
\end{figure}

The consequences of this behaviour for the proper length of the diagonal, as measured from the horizon to the vertex, is shown in Fig. \ref{diagonalfig}. Here we plot the separation of the horizon and the vertex as a fraction of the initial length of a curve in a spatially closed FLRW cosmology that initially subtends the same angle at the centre of the hypersphere. Some similarities with Fig. \ref{horizonsfig} are apparent at early times, with the gradients of the curves corresponding to the six lattices initially having non-zero gradient. The precise lengths of the these curves at $t=0$ are, however, different, and are displayed for reference in Table \ref{scaletable3}, along with the time at which the horizons meet the vertices. It is interesting to note that the time at which the distance between the horizon and the vertex begins to decrease rapidly is also approximately the time at which these curves cross the FLRW curve. The FLRW solution therefore appears to give a good estimate for the time at which the behaviour of the lattice solutions changes, even if it does not give a very good approximation to the functional form of the evolution of the diagonal itself.

\begin{table}[t!]
\begin{center}
\begin{tabular}{|c|c|c|}
\hline
\; {\bf Lattice} \; & \; \bf{ $\left.  \frac{\ell}{ \ell_{{\rm FLRW}}} \right|_{t=0}$} \; & \;  $\frac{t_{\rm collapse}}{n \times m}$    \; \\
\hline
$5-$cell & $1.579$ &  $0.724$\\
$8-$cell & $1.515$ & $0.691$\\
$16-$cell & $1.309$ & $0.723$\\
$24-$cell & $1.311$ & $0.636$\\
$120-$cell & $1.175$ & $0.445$\\
$600-$cell & $1.045$ & $0.234$\\
\hline
\end{tabular}
\end{center}
\caption{{\protect{\textit{The fractional difference in scale between the length of a diagonal in each of the lattices and in a FLRW solution with the same total proper mass at $t=0$, and the time at which the length of the diagonal vanishes.}}}}
\label{scaletable3}
\end{table}

\section{Discussion}

A key assumption in standard cosmological modelling is that of the homogeneity of the Universe. The evidence for the acceleration of the Universe at late times, for example, is almost all obtained by interpreting the observational data within models that assume spatial homogeneity and isotropy from the outset. The fact that this acceleration implies the existence of exotic matter fields, or modifications to gravity, is, however, itself a very surprising result, and therefore calls for the careful consideration of relaxing or ignoring the assumption of homogeneity. This same goal is also called for by the advent of precision cosmology. With these motivations, a great deal of effort has gone into studying the various aspects of this question in recent years \cite{inhomo,inhomo1,inhomo2,inhomo2b,inhomo3,inhomo4,inhomo5,inhomo6,inhomo7,inhomo8}.

Given the enormous complexity of the observed inhomogeneities of the Universe, a useful methodology is to proceed in a step-by-step approach. With this in mind, we have made a detailed study of the exact evolution of a set of cosmological models with discretized matter content. A central tool in our approach has been the employment of symmetries in and about submanifolds of spacetimes that themselves possess no continuous global symmetries. The existence of local rotational symmetry about some curves in the spacetime has allowed us to investigate the evolution of these models by studying the spacetime in their vicinity. Such curves include the edges of the cells that constitute the lattices that we use to build our models, as well as the curves that connect neighbouring masses.

We find that this approach is sufficient to allow us to determine a number of interesting features in the evolution of these models. In particular, we find that while cell vertices and edges remain non-singular throughout the entire evolution of the Universe, other points collapse to anisotropic singularities in finite time, with the regions interior to the MOTS of black holes collapsing first, and with the spacetime in the vicinity of these surfaces often becoming singular before the horizons of neighbouring particles are allowed to merge. Thus, while these models possess regions that behave in a very Friedmann-like way, they also have regions whose evolution deviates radically from {  that normally found in FLRW models.}
In particular, we find that accelerating expansion of cell edges is possible without any violation of the energy conditions, and that we can even have acceleration that would normally require a phantom-like fluid in FLRW cosmology.

In addition to demonstrating the variety of modes of behaviour that are possible in different regions of 
{  the discrete models considered here}, one of the important features of this approach is that it allows their evolution to be studied throughout their entire history. This is far beyond what has so far been possible using any other methods. 
\vspace{10pt}
\begin{center}
\noindent {\bf Acknowledgements}
\end{center}

{TC and RT acknowledge the hospitality of Stockholm University, and KR acknowledges that of the Astronomy Unit at QMUL. RT is supported by STFC grant ST/J001546/1, and TC by a STFC fellowship. DG is an Erasmus Mundus Joint Doctorate IRAP PhD student and is supported by the Erasmus Mundus Joint Doctorate Program by Grant Number 2011-1640 from the EACEA of the European Commission.}




\begin{thebibliography}{tbds}

\bibitem{lightcone} Clarkson, C. \& Maartens, R., {\it Class. Quant. Grav.} {\bf 2}, 124008 (2010).

\bibitem{lattice1} Clifton, T., Rosquist, K. \& Tavakol, R.,
{\it Phys. Rev. D} {\bf 86}, 043506 (2012).

\bibitem{LandW} Lindquist, R. W. \& Wheeler, J. A., {\it Rev. Mod. Phys.} {\bf 29}, 432 (1957); erratum, {\it Rev. Mod. Phys.} {\bf 31}, 839 (1959).

\bibitem{CandF1} Clifton, T. \& Ferreira, P. G., {\it Phys. Rev. D} {\bf 80}, 103503 (2009); erratum {\it Phys. Rev. D} {\bf 84}, 109902 (2011).

\bibitem{CandF2} Clifton, T. \& Ferreira, P. G., {\it JCAP} {\bf 10}, 26 (2009).

\bibitem{CandF3} Clifton, T., Ferreira, P. G. \& O'Donnell, K., {\it Phys. Rev. D} {\bf 85}, 023502 (2012).

\bibitem{Clifton} Clifton, T., {\it Class. Quant. Grav.} {\bf 28}, 164011 (2011).

\bibitem{Larena1} Bruneton, J.-P. \& Larena, J., {\it Class. Quant. Grav.} {\bf 29}, 155001 (2012).

\bibitem{Larena2}  Bruneton, J.-P. \& Larena, J., {\it Class. Quant. Grav.} {\bf 30}, 025002 (2013).

\bibitem{Yoo} Yoo, C. M., Abe, H., Nakao, K. \& Takamori, Y., {\it Phys. Rev. D} {\bf 86}, 044027 (2012).

\bibitem{BandK8cell}  Bentivegna, E. \& Korzy\'{n}ski, M., {\it Class. Quant. Grav.} {\bf 29}, 165007 (2012).

\bibitem{BandK2} Bentivegna, E., arXiv:1305.5576.

\bibitem{BandK3} Bentivegna, E. \& Korzy\'{n}ski, M., {\it Class. Quant. Grav.} {\bf 30}, 235008 (2013).

\bibitem{Misner-63} Misner, C. W., {\it Ann. Phys.} {\bf 24}, 102 (1963).

\bibitem{Brill-Lindquist-63} Brill, D. R. \& Lindquist, R. W., 
{\it Phys. Rev.} {\bf 131}, 471 (1963).

\bibitem{Gibbons} Gibbons, G. W., 
{\it Commun. Math. Phys.} {\bf 27}, 87 (1972).

\bibitem{Cadez} Cadez, A., 
{\it Ann. Phys.} {\bf 83}, 449 (1974).

\bibitem{Newton1} Saari, D. G., {\it Astrophys. J.} {\bf 165}, 399 (1971).

\bibitem{Newton2} Saari, D. G., {\it Trans. Am. Mat. Soc.} {\bf 156}, 219 (1971).

\bibitem{Newton3} Marchal, C. \& Saari, D. G., {\it J. Diff. Eq.} {\bf 20}, 150 (1976).

\bibitem{Newton4} Saari, D. G., {\it Celestial Mechanics} {\bf 21}, 9 (1980).

\bibitem{Newton5} Battye, R. A. \& Gibbons, G. W., {\it Proc. Roy. Soc. Lond.} {\bf A459}, 911 (2003).

\bibitem{Newton6} Gibbons, G. W. \& Patricot, C. E., {\it Class. Quant. Grav.} {\bf 20}, 5225 (2003).

\bibitem{Newton7} Saari, D. G., Collisions, rings, and other Newtonian N-body problems, American Mathematical Society (2005).

\bibitem{Newton8} Gibbons, G. W. \& Ellis, G. F. R., arXiv:1308.1852.

\bibitem{vanElst-Uggla-97} van Elst, H. \& Uggla, C.,
{\it Class. Quant. Grav.} {\bf 14}, 2673 (1997).

\bibitem{Schucking} Ellis, G. F. R. \& MacCallum, M. A. H.,
{\it Comm. Math Phys.} {\bf 12}, 108 (1969).

\bibitem{Ellis-vanElst-1998} Ellis, G. F. R. \& van Elst, H.,
Carg{\`e}se Lectures (1998), gr-qc/9812046.

\bibitem{Coxeter} Coxeter, H. M. S., Regular Polytopes, Methuen and Company
Ltd., London (1948).

\bibitem{Ellis-67} Ellis, G. F. R.,
{\it J. Math. Phys.} {\bf 8}, 1171 (1967).

\bibitem{Stewart-Ellis-68} Stewart, J. \& Ellis G. F. R.,
{\it J. Math. Phys.} {\bf 9}, 1072 (1968).

\bibitem{Bolejko} Bolejko, K., Krasi\'{n}ski, A., Hellaby, C. \& C\'{e}l\'{e}rier, M.-N., Structures in the Universe by Exact Methods, CUP, Cambridge (2009).

\bibitem{korGR} Korzy\'{n}ski, M.,
work presented at the GR20 meeting in Warsaw (2013).

\bibitem{baum} Baumgarte, T. W. \& Shapiro, S. L., Numerical Relativity: Solving Einstein's Equations on the Computer, CUP, Cambridge (2010).

\bibitem{Ash} Ashtekar, A. \& Krishnan, B., {\it Living Rev. Rel.} {\bf 7}, 10 (2004).

\bibitem{Eis} Eisenhart, L. P., Riemannian Geometry, Princeton (1964).

\bibitem{inhomo} Zalaletdinov, R. M., {\it Bull. Astron. Soc. India} {\bf 25}, 401 (1997).

\bibitem{inhomo1} Buchert, T., {\it Gen. Rel. Grav.} {\bf 40}, 467 (2008).

\bibitem{inhomo2} van den Hoogen, R., arXiv:1003.4020.

\bibitem{inhomo2b} Andersson, L. \& Coley, A., {\it Class. Quant. Grav.} {\bf 28}, 160301 (2011).

\bibitem{inhomo3} Ellis, G. F. R, {\it Class. Quant. Grav.} {\bf 28}, 164001 (2011).

\bibitem{inhomo4} Wiltshire, D., {\it Class. Quant. Grav.} {\bf 28}, 164006 (2011).

\bibitem{inhomo5} Clarkson, C., Ellis, G. F. R., Larena, J. \& Umeh, O., {\it Rept. Prog. Phys.} {\bf 74}, 112901 (2011).

\bibitem{inhomo6} Buchert, T. \& R\"{a}s\"{a}nen, S., {\it Ann. Rev. Nucl. Part. Sci.} {\bf 62}, 57 (2012).

\bibitem{inhomo7} Clarkson, C., {\it Comptes Rendus Physique} {\bf 13}, 682 (2012).

\bibitem{inhomo8} Clifton, T., {\it Int. J. Mod. Phys. D} {\bf 22}, 1330004 (2013).


\end{thebibliography}
\end{document}